\documentclass{aastex}
\usepackage{emulateapj5}

\newcommand{\simgt}{\lower 2pt \hbox{$\, \buildrel {\scriptstyle >}\over {\scriptstyle\sim}\,$}}
\newcommand{\simlt}{\lower 2pt \hbox{$\, \buildrel {\scriptstyle <}\over {\scriptstyle\sim}\,$}}

\newcommand{\rxj}{RX~J1131$-$1231}

\newcommand{\he}{HE~1104$-$1805}

\newcommand{\chandra}{{\emph{Chandra}}}

\slugcomment{Received 2008 May 24}
\shorttitle{X-ray Microlensing}
\shortauthors{CHARTAS ET AL.}

\begin{document}

\def\sarc{$^{\prime\prime}\!\!.$}
\def\arcsec{$^{\prime\prime}$}
\def\beginrefer{\section*{References}%
\begin{quotation}\mbox{}\par}
\def\refer#1\par{{\setlength{\parindent}{-\leftmargin}\indent#1\par}}
\def\endrefer{\end{quotation}}
\title{X-ray Microlensing in RXJ1131--1231 and HE1104--1805}

\author{G. Chartas\altaffilmark{1}, C. S. Kochanek\altaffilmark{2}, X. Dai\altaffilmark{2}, S. Poindexter\altaffilmark{2}, \& G. Garmire\altaffilmark{1}}

\altaffiltext{1}{Department of Astronomy \& Astrophysics, Pennsylvania State University,
University Park, PA 16802, chartas@astro.psu.edu}

\altaffiltext{2}{Department of Astronomy and the Center for Cosmology and Astroparticle Physics, The Ohio State
University, Columbus, OH 43210, USA}

\begin{abstract}

We present results from a monitoring campaign performed with the 
{\it Chandra X-ray Observatory} of the gravitationally lensed quasars RX~J1131$-$1231
and HE~1104$-$1805. We detect significant X-ray variability in all images of both quasars.
The flux variability detected in image A of RX~J1131$-$1231 is of particular interest because of its 
high amplitude (a factor of $\sim$ 20). We interpret it as arising from microlensing since the variability is uncorrelated
with that of the other images and the X-ray flux ratios show larger changes than
the optical as we would expect for microlensing of the more compact X-ray emission
regions. The differences between the X-ray and optical flux ratios of HE~1104$-$1805 are less dramatic, but there 
is no significant soft X-ray or dust absorption, implying the presence of X-ray microlensing in this system as well. Combining the X-ray data with the optical light curves
we find that the X-ray emitting region of HE~1104$-$1805 is compact with a half-light radius $\simlt 6 r_g$,
where the gravitational radius is r$_{\rm g}$ = 3.6 $\times$ 10$^{14}$~cm, 
thus placing significant constraints on AGN corona models.
We also find that the microlensing in HE~1104$-$1805
favors mass models for the lens galaxy that are dominated
by dark matter. Finally, we better characterize the massive foreground cluster near RX~J1131$-$1231, 
set limits on other sources of extended
X-ray emission, and limit the fluxes of any central odd images to be 30--50 (3$\sigma$) times fainter than
the observed images.

\end{abstract}


\section{INTRODUCTION}

Resolving the emission regions of distant quasars is beyond the current capabilities of 
present-day telescopes, as direct imaging of accretion disks requires angular resolutions on 
the order of tens of nanoarcseconds at $z$  $\sim$ 1. Until the spatial resolution of telescopes
reaches this limit, we will have to rely on indirect methods of mapping the emission regions of quasars. 
Such methods include light-travel time arguments, reverberation mapping of the
broad line region (Blandford \& McKee 1982; Peterson 1993, Netzer \& Peterson 1997),
reverberation mapping of the Fe K$\alpha$ emission region (Young \& Reynolds 2000), 
occultation measurements of the central X-ray source by orbiting Compton thick clouds
(Risaliti et al, 2007), and microlensing of the continuum and line emission regions
(e.g., Grieger et al. 1988 and 1991; Schneider, Ehlers \& Falco 1992; Gould \& Gaudi 1997; 
Agol \& Krolik 1999; Yonehara et al. 1999; Mineshige \& Yonehara 1999; Chartas et al. 2002a; Popovic et al. 2003; Blackburne et al. 2006; Pooley et al. 2006, 2007; Kochanek et al. 2007; Jovanovic et al. 2008; Morgan et al. 2008a, 2008b).

We observe microlensing as uncorrelated variability in the brightnesses of the images
of a lensed quasar, where the amplitude of the variability is determined by the 
size of the emission regions (e.g., Lewis et al. 1998; Popovic \& Chartas 2005).
The largest components, the radio, dust and optical line emission regions, should show
little or no microlensing variability.  The thermal continuum emission from the 
disk should show greater variability at shorter wavelengths corresponding to
smaller disk radii and higher temperatures, and this has been observed by
Poindexter et al. (2008) and Anguita et al. (2008).  The non-thermal X-ray
emission is thought to be dominated by inverse Compton scattering of UV photons
from the disk (e.g., Reynolds \& Nowak 2003), some of which is reprocessed into Fe~K$\alpha$ line emission 
(e.g., George \& Fabian 1991; Mushotzky et al. 1993)
but the geometry and scale of these emission regions is not well understood.

The effects of microlensing on the X-ray fluxes of lensed images have been
reported for many lenses at this point, including Q0957+561 (e.g., Chartas et al. 1995), 
RX J0911+0511 (e.g., Morgan et al. 2001; Chartas et al. 2001), H1413+117 (e.g., Chartas et al. 2004, 2007), 
PG~1115+080 (Pooley et al. 2006; Morgan et al. 2008b), and \rxj\ (e.g., Blackburne et al. 2006;
Kochanek et al. 2007; Dai et al. 2008).  The differences between the X-ray flux
ratios and optical flux ratios can be used to estimate the source size after accounting
for absorption by the interstellar medium of the lens
(e.g., Dai \& Kochanek 2008) and the effects of substructures (satellites) in
the lens halo (Mao \& Schneider 1998).  Pooley et al. (2007) used observed
X-ray and optical flux ratios for ten four-image lensed quasars to show that
the X-ray source is much more strongly microlensed than the optical source,
indicating that the X-ray emission region is more compact than the optical.
They also found that their estimates of the optical source
sizes were significantly larger than would be expected based on the observed,
magnification-corrected optical fluxes.  Morgan et al. (2008a) measured the
sizes of the optical emission regions of 11 quasars using microlensing to
find that the sizes scaled with black hole mass as expected from thin disk
theory (Shakura \& Sunyaev 1973) and had approximately the sizes expected
from thin disk theory, but confirmed the Pooley et al. (2007) result that these sizes, both microlensing and
theoretical, are larger than expected given the observed optical flux and the
same disk model.  

Accurate microlensing estimates of the sizes of the X-ray emission regions requires 
X-ray light curves rather than isolated epochs.  First, the absolute 
magnifications of lensed images are not well understood because 
substructure in the lens galaxy can modify the fluxes from the 
predictions of lens models (e.g., Mao \& Schneider 1998; Dalal \& Kochanek 2002).  While comparing optical and X-ray flux
ratios provides estimates for the difference in size between the 
emission regions, the lack of a secure estimate of the intrinsic
flux ratios makes it difficult to set an absolute scale.  Second,
quasars have intrinsic variability that appears in the images
with relative time delays, which means that instantaneous flux ratios
are contaminated by the effects of time variability modulated by the
delays.  Third, while the source size ultimately determines the 
amplitude of microlensing variability, the instantaneous values are
a combination of the source size and the location of the source
in the complex microlensing magnification patterns.  Monitoring the
variability minimizes these problems and should lead to far more accurate
estimates of the sizes.

We have been monitoring roughly 25 lenses in the optical to measure time 
delays (e.g., Kochanek et al. 2006) and to study quasar structure (e.g.,
Morgan et al. 2008a; Poindexter et al. 2008).  From this sample we
selected a small subset with reasonable X-ray fluxes for monitoring 
with the {\it Chandra X-ray Observatory}.  Given a well-sampled 
optical light curve that can be used to determine the size of the
optical/UV emission region and act as a  microlensing variability ``reference,''
we can afford to sample the X-ray microlensing variability relatively
sparsely.  In this paper we present the X-ray observations of the lensed quasars 
\rxj\ ($z_{\rm s} = 0.658$) and \he\ ($z_{\rm s} = 2.32)$.  
The optical data for \rxj\ and HE~1104--1805 are presented in Morgan et al. (2006) and
Poindexter et al. (2007) respectively, and the optical microlensing results  
are presented in Morgan et al. (2008a) and Poindexter et al. (2008). 
Here we carry out a microlensing analysis of \he, while a companion paper, Dai et al. (2009), does so for \rxj.
We discuss the observations and data analysis in \S2, study the microlensing
of HE~1104--1805 in \S3 and summarize our results in \S4. 
Throughout this paper we adopt a $\Lambda$-dominated cosmology with 
$H_{0}$ = 70~km~s$^{-1}$~Mpc$^{-1}$, 
$\Omega_{\rm \Lambda}$ = 0.7, and  $\Omega_{\rm M}$ = 0.3.

\section{X-RAY OBSERVATIONS AND DATA ANALYSIS}

We monitored \rxj\ and \he\ with the Advanced CCD imaging 
Spectrometer (ACIS; Garmire et al. 2003) on board the {\it Chandra X-ray Observatory} 
(hereafter \chandra) using short ($\sim$ 5~ksec) snapshot observations.
We obtained 5 epochs for each lens in 2006
placing the sources on the back-illuminated S3 chip of ACIS.  We combined these with 
archival observations of \rxj\ in April 2004 and \he\ in February 2000.
A log of the observations that includes observation dates, observed count rates, 
total exposure times, and observation identification numbers is presented in Table 1.
We analyzed the data using the standard software CIAO 4.0 provided by the CXC. 
We used standard CXC threads to screen the data for 
status, grade, and time intervals of acceptable aspect solution and background levels.
We removed the {$\pm$~0\sarc25} spatial
randomization applied to the event positions by the standard pipeline
and instead used the sub-pixel resolution techniques developed by
Tsunemi et al. (2001) and Mori et al. (2001) in order to improve
the image resolution.

The \chandra\ spectra of \rxj\ and \he\ were fit with a variety of models
employing \verb+XSPEC+ version 12 (Arnaud 1996).
For all spectral models of \rxj\ and \he\ we 
included Galactic absorption 
due to neutral gas (Dicky \& Lockman 1990) with column densities of 
 $N_{\rm H}$= 3.6 $\times$ 10$^{20}$~cm$^{-2}$ and 4.6 $\times$ 10$^{20}$~cm$^{-2}$, respectively.
We use 90\% confidence level uncertainties unless otherwise stated.

\subsection{Spatial and Spectral Analysis}

For estimating the X-ray counts of images B, C, and D of \rxj\ 
we extracted events from circular regions with radii of 1.5~arcsec
slightly off-center from the images to reduce contamination from nearby images 
(see panel (c) of Fig.~1). 
To estimate the X-ray counts of image A of \rxj,  which lies between images B and C,
we extracted events within a 0.75~arcsec circular region
centered on A.
Aperture corrections were applied to all images to account for counts not included in the extraction regions.
For estimating the X-ray counts of images A and B of \he\ we extracted 
events from circular regions centered on the images with radii of 
1.5~arcsec. The backgrounds for \rxj\ and \he\ were determined by extracting events within an
annulus centered on the mean location of the images
with inner and outer radii of 7.5~arcsec and
50~arcsec, respectively. 
Contamination between images was minimized and made insignificant by the application of the 
sub-pixel resolution technique and the use of appropriately sized and off-center extraction apertures.
We also corrected for ``pile-up," an instrumental effect that occurs
when two or more X-ray photons strike individual or neighboring CCD pixels within 
one frame time.  Neglecting these corrections can cause spectral distortion, grade migration and 
distortion of the image PSF. We used the  forward spectral-fitting tool LYNX
(Chartas et al. 2000) to estimate the fraction of events lost due to the pile-up
effect.

Table 1 summarizes the observed 0.2--10~keV band source counts.  The pile-up corrections needed to
correctly estimate these counts can be significant.  For example the
counts for images B and C in the 2004 April 12 observation of
\rxj\ have pile-up corrections of 37\% and 16\% respectively.
In their analysis of the 2004 observation of \rxj, Blackburne
et al. (2006) did not include pile-up corrections, so our present
results should be more reliable.  In particular, pile-up usually leads to
energy spectra that appear harder, and this likely explains the differences in the X-ray spectral hardness ratios reported by Blackburne et al. (2006) and our present results.

In Figure 1 we show the Lucy-Richardson deconvolved images in the 0.2--8~keV bandpass of the
\chandra\ observations of \rxj\ and \he.
To estimate the relative X-ray image positions of \rxj\ and \he\ we modeled the \chandra\ images 
using point-spread functions (PSFs) generated by the simulation tool \verb+MARX+ (Wise et al 1997).
The X-ray event locations were binned with a bin-size of 0\sarc0246 to sample the PSF 
sufficiently (an ACIS pixel subtends 0\sarc491). The simulated PSFs were fitted to the \chandra\ data by minimizing the
$C$-statistic formed between the observed and simulated images.
In Table 2 we compare the X-ray image separations to
the observed NICMOS values reported by Morgan et al. (2006) for \rxj\
and by L\'ehar et al. (2000) for \he.
We conclude that the X-ray and optical image positions are consistent given the estimated uncertainties.

The deconvolved images of \rxj\ and \he\  do not show any additional lensed images.
To obtain quantitative limits on possible additional images located at the centers of  \rxj\ and \he\
we  extracted the 0.2--10~keV counts within 0.4~arcsec circles centered 
on the images at the mean lens position of each quasar.
The backgrounds in the central source extraction regions are dominated by the contamination
from the images -- the instrumental and cosmic backgrounds in the 0.4 arcsec extraction circles 
are significantly lower than the contamination.
To estimate the fractional contamination per image we used the 0.2--10~keV counts in 
0.4~arcsec circular apertures placed 1 and 2~arcsec North of image B in \rxj\ and 1 and 2~arcsec Northwest of image A in \he.
The detected 0.2--10~keV counts in the central 0.4 arcsec apertures are 
consistent with the estimated background and contamination from the bright images.

By combining all the data for each lens we set 3$\sigma$ upper limits on the 0.2--10~keV flux
of any central, odd image of $1\times 10^{-14}$ and $4\times 10^{-15}$~ergs~s$^{-1}$~cm$^{-2}$ 
for \rxj\ and \he, respectively.   These limits corresponded to a 3$\sigma$ limit on the
flux ratio of a factor of 50 relative to image C in \rxj, and a factor of 30 relative to image A in \he. While relatively tight, these limits are not strong enough to
constrain the central surface density of the lens (Keeton 2003).  For example, in the models we
use for \he\ below, the expected flux ratio between image A and any central image
is $>10^3$.

We performed fits to the individual spectra of the images of \rxj\ and \he\ using events 
in the 0.4--8~keV energy range with a
model that consisted of a simple power law modified by Galactic absorption.
Due to the moderate S/N of the spectra, we performed these fits using the Cash statistic
which does not require binning of the data, although if we instead use $\chi^2$
statistics we find similar results.  The best-fit parameters of these fits and the 
unabsorbed 0.2--2~keV and 2-10~keV fluxes are presented in Tables  2 and 3.
We found no significant variability (within the 90\% errors) of the photon indices 
(${\Gamma}$) of the spectra of the images of \rxj\ and \he\ with the exception of the 
2004 observation of \rxj\ that showed significant differences of ${\Gamma}$ between images
A/B/C with $\Gamma \simeq 1.44$ and image D with $\Gamma \simeq 1.95$. 
We also considered models that included neutral absorption at the redshifts of the sources or lenses
of \rxj\ and \he. We do not detect any additional neutral absorption at these redshifts.  This
is consistent with the marginal detection of absorption in \he\ in our earlier analysis
(Dai et al. 2006).

In Figure~2 we show the {\it Chandra} image of the 2004 observation of the lensed system \rxj\ and the surrounding field.
To reduce background contamination and to enhance possible soft extended X-ray emission we filtered the image to include 
only photons with energies ranging between 0.4 and 3.0~keV. 
The image was binned with a bin size of 0.5 arcsec and adaptively smoothed with the tool
CSMOOTH developed by Ebeling et al. (2000).
CSMOOTH smooths a two-dimensional image with a circular Gaussian kernal of varying radius.
We confirm the extended soft X-ray emission reported by Morgan et al. (2006)
centered at 11 32 1.6, $-$12 31 6.5 (J2000) and about 158~arcsec Northeast of image D of \rxj.
This emission comes from a foreground cluster at $z=0.1$.
Morgan et al. (2006) also reported a possible detection at a low significance level 
(3~$\sigma$) of extended emission centered
33~arcsec Southwest of \rxj\ at 11 31 50.1, $-$12 32 23 (J2000).
We combined all six observations of \rxj\ listed in Table~1,
for a total exposure time of 33.8~ksec, to better limit this possible source.  All the observations 
from 2006 were performed in sub-array mode to mitigate pile-up and with 
the reduced field of view they do not include the foreground cluster to the
Northeast of the lens.  Figure~3 shows the adaptively smoothed image including only 
photons with energies from 0.4 to 3.0~keV.  We find no significant sources
of extended emission near the lens beyond the foreground cluster.  
The 0.2--2.~keV $3\sigma$ upper flux and luminosity
limits are $5 \times 10^{-15}$~ergs~cm$^{-2}$~s$^{-1}$ and 
$1.3 \times 10^{42}$~ergs~s$^{-1}$, respectively, assuming a cluster with a 
temperature of $1.5$~keV at the lens redshift with
an abundance of $0.3$ solar and an extraction radius of 30~arcsec.    

We fit the emission from the foreground cluster
using a $\beta$ model 
for the cluster brightness profile combined with a uniform background 
of 0.005 events per pixel.  Prior to performing the fit
we binned the image in 1{\arcsec} pixels and smoothed this with a Gaussian ($\sigma$ = 3\arcsec).
The fits were performed with the CXC software package \verb+SHERPA+. 
We find that the cluster center is ${\Delta}{\alpha}$ = 150{\sarc}2 $\pm$ 0\sarc5 East 
and ${\Delta}{\delta}$ = 52{\sarc}3 $\pm$ 0\sarc5 North of image D.
The smoothed intensity distribution is nearly round
with an ellipticity of ${\epsilon} = $ 0.10 $\pm$ 0.3.
The best-fit values for $\beta$ and the core radius of the cluster
are $\beta$ = 0.4 $\pm$ 0.2 and r$_{0}$ = 4{\sarc}2$_{-0.3}^{+0.4}$ (8~kpc),
respectively. 

We extracted the spectrum of this $z =0.1$ cluster 
from a 50 arcsec radius circle centered on the X-ray cluster center.
This spectrum was fit with a simple 
model consisting of an emission spectrum from hot diffuse gas based 
on the XSPEC model mekal modified by Galactic absorption. 
We obtain best-fit values for the temperature and metal abundances
of 1.2$^{+0.2}_{-0.1}$~keV and A = 0.5$^{+0.5}_{-0.2}$,
respectively (both errors are at the 90\% confidence level).
The 2--10~keV luminosity of this cluster of galaxies is 1.7 $\times$ 10$^{42}$~ergs~sec$^{-1}$.
These are consistent with the Morgan et al. (2006) estimates and the
values for the temperature, $\beta$ and the core radius are consistent
with the observed correlations of clusters (e.g. Jones \& Forman 1999).

We also stacked the images of the six observations of \he\
listed in Table 1 for a total exposure time of 71.8~ksec.
Figure~4 shows the adaptively smoothed image of \he\ including only 
photons with energies from 0.4 to 3.0~keV.  We find no significant sources
of extended emission. The 0.2--2~keV $3\sigma$ upper flux and luminosity
limits are $3.4 \times 10^{-15}$~ergs~cm$^{-2}$~s$^{-1}$ and 
$5.7 \times 10^{42}$~ergs~s$^{-1}$, respectively, assuming a cluster with a 
temperature of $1.5$~keV at the lens redshift with
an abundance of $0.3$ solar and an extraction radius of 30~arcsec.

\subsection{Timing Analysis}

The 0.2--10~keV time-delay corrected light curves of the images of \rxj\ and \he\ are shown in 
Figures 5a and 5b respectively.  Significant X-ray flux variability is detected in all images.
\rxj\ shows a mixture of intrinsic and microlensing variability.  If we normalize
the first epoch for image D to match the optical light curves, we find that images
B, C and D roughly track the optical light curves and show similarities in
their time variability.  
In Figure 6a we show the flux ratios F$_{\rm A}$/F$_{\rm C}$, F$_{\rm B}$/F$_{\rm C}$, and F$_{\rm D}$/F$_{\rm C}$
in the 0.2--10 keV and $R$-bands of \rxj.
The flux ratio F$_{\rm B}$/F$_{\rm C}$ is almost constant in both the X-ray and optical bands indicating that 
the observed variability in images B and C is correlated and therefore mostly intrinsic in origin.
Image A shows a completely different behavior. It is 
a factor of $\sim$ 4 fainter in the X-rays than the optical relative to C/D in 2004, comparable to the optical in the spring of 2006, and
a factor of $\sim$ 5 brighter by the fall of 2006. Since the 2006
observations come in clusters with temporal separations of order the A/B/C
time delays, we can be confident that these differences are not due to 
intrinsic variability modulated by the time delay. 
The significant rise of the F$_{\rm A}$/F$_{\rm C}$ ratio in the X-ray band 
compared to the one measured in the $R$-band supports our microlensing 
interpretation of the flux enhancement in image A of \rxj. 
We note that the flux ratios reported in our analysis for the 2004 observation of \rxj\ 
are different from those reported by Blackburne et al. (2006)
because of the correction for the loss of counts 
due to pile-up and the correction for the spectral distortion
due to pile-up.
We see no strong evidence for variations in the spectral hardness, but this is a 
weak statement given the limited count rates. 

\he is also affected by intrinsic X-ray variability. In particular, the peak
in the optical flux near 4100~days seems likely to be associated with a peak in
the X-ray flux. In Figure 6b we show the flux ratios F$_{\rm A}$/F$_{\rm B}$ 
in the 0.2--10~keV and $R$-bands of \he. 
The February/March and October/November epochs are 
roughly separated by the time delay, so we can reasonably conclude that the 
X-ray flux ratio near 4050~days (observed A) 
is approximately F$_{\rm A}$/F$_{\rm B}$$\simeq 1$ 
and very different from the optical flux ratio of F$_{\rm A}$/F$_{\rm B}$$\simeq 3$.  Since 
there is no evidence for significant soft X-ray or dust absorption (see \S2.3,
Dai et al. 2006, Poindexter et al. 2008)  this difference in flux ratios
must be due to microlensing.  
In \S 3 we proceed with fitting the light-curves of HE 1104-1805
with a sophisticated microlensing model. The microlensing analysis of the light-curves of RX J 1131-1231 
will be presented in a companion paper by Dai et al. (2009).
As with \rxj, we see no evidence for variations
in the spectral hardness, largely due to the large uncertainties.



\section{X-ray Microlensing in HE~1104$-$1805}

We modeled the microlensing in \he\ as in Morgan et al. (2008a) and Poindexter et al. (2008)
using a fixed 162~day time delay, the $R$-band light curve constructed from our SMARTS data (Poindexter et al. 2007), OGLE (Wyrzykowski et al. 2003), and Ofek \& Maoz (2003), and the delay-corrected X-ray flux measurement.
The main free parameters of our lensing model for \he\
are the X-ray and UV (rest-frame) scale lengths, a microlens mass scale,
a mass fraction of the local surface density comprised of stars,
and a velocity vector describing the motion of the AGN emission region
across the microlensing caustics.
We used the lens model sequence from Poindexter et al. (2007), where we start from a constant mass-to-light
ratio model for the mass distribution, defined by $f_{M/L}=1$, and then reduce its mass in 10\% increments
while adding an NFW halo, where $f_{M/L}=0$ would correspond to a pure halo model.  The measured time
delay requires $f_{M/L} \simeq 0.3$.  We made 8 random realizations of the star fields near each
image and then generated $8192^{2}$ pixel magnification maps using the methods of Kochanek (2004).  The maps
had an outer scale of $10 R_{\rm E} = 4.3 \times 10^{16}$~cm and an inner scale of $10 R_{E}/8192 = 4.3\times 10^{13}$~cm.
Based on its emission line widths, Peng et al. (2006) estimated that the black hole mass in \he\
is $M_{\rm BH} = 2.4 \times 10^{9} M_\odot$, corresponding to a gravitational radius of $r_{\rm g}=GM_{\rm BH}/c^{2}=3.6 \times 10^{14}$~cm
that is well-resolved in the magnification maps.   We modeled the surface brightness of the emission
regions as a face-on, thin disk (Shakura \& Sunyaev 1973) without the central temperature depression,
$$   f_\nu = { 2 h_{\rm p} c \over \lambda_{rest}^3 } \left[ \exp\left( { R \over R_{\lambda_{\rm rest}}}\right) -1 \right]^{-1}
   \label{eqn:fnu}
$$
where the scale length 
$$ R_{\lambda_{rest}} =
     \left[  45 G \lambda_{\rm rest}^{4} M_{\rm BH} \dot{M} \over 16 \pi^{6} h_{\rm p} c^{2} \right]^{1/3}
$$

$$     
{}      = 9.7 \times 10^{15} \left({\lambda_{\rm rest}\over \mu\hbox{m}}\right)^{4/3}
        \left( { M_{\rm BH} \over 10^{9} M_\odot}\right)^{2/3}
        \left( { L \over \eta L_{\rm E} } \right)^{1/3}~\hbox{cm}
$$ 
is the radius at which the disk temperature matches the rest wavelength of the observations, 
$kT=h_{\rm p}c/\lambda_{\rm rest}$, $\dot{M}$ is the accretion rate, $L/L_{\rm E}$ is the luminosity in
units of the Eddington luminosity and $\eta=L/(\dot{M}c^2)$ is the accretion efficiency.
We scale the results assuming a radiative efficiency of $\eta=10\%$.  While the available
energy of accretion is larger, the division of that energy between radiative, advective and 
kinetic losses is not directly calculable at present. Studies of the growth of black holes
tend to find that the average radiative efficiency of accretion is less than 10\% 
(e.g. Shankar et al. 2007), and low radiative efficiencies are also needed
to reconcile thin disk theory with microlensing accretion disk size estimates (e.g., Morgan et al. 2007).  

For the UV (rest-frame) emission we can neglect the central temperature depression provided 
$R_{\lambda_{\rm rest}} >> R_{in}$ where $R_{in} \simeq 2 r_g$ is the inner edge of the disk.
While this thermal accretion disk model is not appropriate for the non-thermal X-ray emission,
Mortonson et al. (2005) have shown that microlensing essentially measures the half-light
radius of the emission region, so we used the same profile for the X-rays to simplify 
the computations but regard it as a measurement of the half-light radius 
$R_{1/2}= 2.44 R_\lambda$.  The microlenses were given a power-law mass distribution
$dN/dM \propto M^{-1.3}$ with a dynamic range in mass of a factor of 50, roughly
matching the Galactic disk mass function of Gould (2000), and we use a uniform 
prior on the mean microlens mass of $0.1M_\odot \leq \langle M\rangle \leq M_\odot$.
We modeled the probability distribution for the physical velocities as in Poindexter
et al. (2008) where our projected velocity onto the lens plane is $73$~km/s, the 
velocity dispersion of the lens is $308$~km/s based on the critical radius of the
mass model, and the lens and source have rms peculiar velocities of $153$ and
$71$~km/s respectively.
The data were then modeled using the Bayesian Monte Carlo method of Kochanek (2004).

The microlensing data constrains the source velocity and size in ``Einstein'' length
units of $\langle M\rangle^{1/2}$~cm.  We can determine the true sizes using the
prior on the microlens masses, the prior on the physical velocities or both.  Figure 7
shows the probability distribution found for $\langle M\rangle$ assuming the 
prior on the physical velocities.  
While the distribution peaks at somewhat lower mass ($\sim 0.05 M_\odot$) than the 
range of our prior ($0.1M_\odot$ to $1.0M_\odot$),
the distribution is so broad ($0.005M_\odot$ to $0.9M_\odot$ at
1-$\sigma$) that we can view them as statistically consistent.
For example, none of the results depend significantly on the
inclusion of the mass prior.
Since the mass scales as $\langle M \rangle \propto v_e^2/\hat{v}_e^2$ where $v_e$ is the physical
velocity from the prior and $\hat{v}_e$ is the Einstein-unit velocity estimated
from the microlensing models, it is difficult to tightly constrain the mean mass with a single
lens given the unknown peculiar velocities of the lens and source even with perfect
knowledge of the Einstein-unit velocity.   

Figure 8 shows the raw estimates of the source sizes $\hat{r}_s$ in Einstein units,
where $r_s=\hat{r}_s \langle M/M_\odot\rangle^{1/2}$.  Both of these are scaled source sizes $R_\lambda$
defined by the emission profile presented earlier in this section. 
The consequence of
adding the prior on the microlens masses is to drive the source sizes $\hat{r}_s$ to be modestly
smaller.  In essence, a given level of microlensing variability can be created by
a large source moving rapidly (big $\hat{r}_s$ and $\hat{v}_e$) or a small source
moving more slowly (small $\hat{r}_s$ and $\hat{v}_e$).  But high effective velocities
correspond to small mass scales given a fixed physical velocity, so imposing the
mass prior given the distribution seen in Figure 7 will favor lower
velocities $\hat{v}_e$ and smaller sources $\hat{r}_s$.  It will have less effect
on the physical source size 
$r_s=\hat{r}_s \langle M/M_\odot\rangle^{1/2}$
because the smaller size $\hat{r}_s$ is balanced by the larger mass scale $\langle M/M_\odot\rangle$ 
(see Kochanek 2004).  While Figure 8 truncates the probability distribution for
the X-ray source at the pixel scale of the magnification patterns, it would not converge
to zero if we used magnification patterns with smaller pixel scales to extend the distribution
to smaller source sizes. This is a consequence of not having a well-sampled X-ray light
curve.  The source must be small enough to allow the X-ray flux ratio to differ from
the optical, but still smaller sources are disfavored because even more discrepant
flux ratios become more likely.  However, reducing the size still further has no effect
on the probabilities because it converges to the likelihood of the data points avoiding
precise caustic crossings.  Unfortunately, this does mean that we will obtain robust
upper limits on the X-ray source size but not lower limits.

Figure 9 shows the results for the rest-frame UV and X-ray source sizes
$R_\lambda$ assuming either uniform or logarithmic priors for the size. We used $R_\lambda$
for both so that the sizes are easily compared.  If we regard the source size at 2000\AA\ as 
measuring a projected area, then $R_\lambda$ scales as $(\cos i)^{-1/2}$ for inclination angle $i$
of a thin inclined accretion disk.  
Similarly, the X-ray size in Figure 9 
should be increased by a factor of
$2.44$ to convert from $R_\lambda$ to the half-light radius $R_{1/2}$.  The disk size at 
2000\AA\ ($R$-band in the rest frame at $z_s=2.32$) is 
$15.6 \leq \log(R_{200}\sqrt{\cos (i)}/\hbox{cm}) \leq 16.2$ for a logarithmic prior on the size
and 
$15.8 \leq \log(R_{200}\sqrt{\cos (i)}/\hbox{cm}) \leq 16.3$ for a linear prior on the size.
This agrees well with our earlier estimates in Poindexter et al. (2008) and Morgan
et al. (2008a).  The X-ray size is less certain and more prior-dependent, with an
estimated half-light radius
$14.2 \leq \log(R_{X,1/2}/\hbox{cm}) \leq 15.0$ for a logarithmic prior on the size
and
$14.6 \leq \log(R_{X,1/2}/\hbox{cm}) \leq 15.3$ for the linear prior on the
size.  We note that the difference in the sizes is entirely driven by the
significant differences between the X-ray and optical flux ratios at the
time of the X-ray observation.  

The microlensing data also prefer lens models with low stellar mass fractions,
as shown in Figure 10.
The probability distribution peaks at
$f_{M/L} \simeq 0.2$, in relatively good agreement with the value of $f_{M/L}=0.3$
that agrees best with the measured time delay and a Hubble constant of 
$H_0=72$~km~s$^{-1}$~Mpc$^{-1}$.   While this preliminary result does not
lead to a microlensing constraint on $f_{M/L}$ that is tight enough to 
break the degeneracies between the radial mass distribution and the Hubble
constant (see Kochanek 2002), it is encouraging that there is agreement with the time delays
given the expected value of $H_0$ and that the microlensing requires a dark
matter-dominated mass distribution.
A detailed microlensing analysis of \rxj\ that provides similar constraints on the 
structure of the lens and the structure of the quasar is presented in Dai et al. (2009). 

Figures 11 and 12 show an example of a trial that
fits the data well.  In this trial the Einstein unit source sizes are 
$\log \hat{R}_\lambda =14.2$ and $16.2$ for the X-ray and UV rest-frame sources
respectively and the effective velocity corresponds to a sensible microlens
mass scale.  We chose this solution mainly because of the curious origin
of the initial $R$-band peak, where what might be interpreted as a single
heavily smoothed caustic crossing is actually a pair 
of competing caustic crossings in the A and B images, where the A
image dominates because of its shallower magnification gradients. The
X-ray epoch then corresponds to time when the source is in a valley
between caustics in both images.  We note that this particular solution
probably would not reproduce the wavelength dependence of the optical
and IR flux ratios in HE~1104$-$1805 discussed and modeled Poindexter et al. (2008).

\section{DISCUSSION AND CONCLUSIONS}

While there is a consensus that the origin of the keV X-ray continuum
emission from quasars is inverse Compton scattering (e.g., see the review
of Reynolds \& Nowak 2003), the extent and geometric configuration of
the emission region is less clear.  In the standard disk-corona model
(e.g., Haardt \& Maraschi 1991; Merloni 2003), it is produced in a hot,
extended corona surrounding the disk.  
In the general relativistic MHD simulations of 
Hirose et al. (2004, also Machida \& Matsumoto 2003) the region of the corona 
with the highest current densities lies close to the inner edge of the disk.
Under the assumption that these are also the regions with the highest
dissipation rates for heating the corona (see Hirose et al. 2004), they
will dominate the X-ray emission and we would expect the X-ray source
size to track the inner edge of the disk.

Microlensing of the X-ray source allows us to observationally measure
the size of the X-ray emission region.  We start with well-sampled
optical light curves that allow us to determine the size of the
accretion disk.  Morgan et al. (2008a) found a clear scaling of the disk 
size with black hole mass that is consistent with the $M_{BH}^{2/3}$ scaling 
of simple thin disk theory for a fixed Eddington accretion rate (Shakura \& Sunyaev 1973). 
We can estimate the disk size by two routes other than microlensing.
First we can estimate it from the observed flux at some wavelength.
Size estimates from the observed flux do not depend on the
black hole mass but depend only on the assumed temperature profile
and the assumption that the disk radiates as a black body at that temperature.
Second, thin disk theory predicts the scale length through the expression
we presented in the beginning of \S 3. Size estimates from thin disk theory
depend on the black-hole mass, the Eddington factor and the wavelength.
The disk size from our microlensing analysis is roughly consistent with thin disk theory if the accretion efficiency
is relatively low, but both the microlensing and theoretical sizes are significantly
larger than that estimated from the observed optical flux of the quasar and the
same disk model.  This mismatch between the sizes estimated from microlensing and
from the flux was first pointed out by Pooley et al. (2007).  One solution
to this problem may be to use a shallower temperature profile than the $T \propto R^{-3/4}$
of simple thin disk theory, and Poindexter et al. (2008) found some evidence for
a shallower profile in \he\ by examining the optical/near-IR wavelength dependence
of microlensing in this system.

Our X-ray monitoring observations of \rxj\ and \he\ show 
significant flux variability in all images of both quasars. 
The factor of $\sim 20$ X-ray flux variability in image A of \rxj\ is
interesting not only because of the high amplitude but also because it
is a highly magnified saddle point image.  Schechter \& Wambsganss
(2002) pointed out that in a dark matter dominated galaxy, microlensing
is in a regime where images at minima and saddle points of the time-delay 
surface behave differently. In particular, the saddle point
images should show significantly higher variability amplitudes, just
as we observe in \rxj.  The relatively large sizes of accretion disks
in the optical relative to the typical Einstein radius of the stars
seems to have masked much of this effect in the optical signatures
of microlensing -- for example, where the X-ray flux ratio $F_A/F_C$
changed by a factor of 4 during our monitoring period, the optical
flux ratio changed by only a factor of 1.4.  The differences between
the X-ray and optical flux ratios of \he\ also imply the
presence of X-ray microlensing.  Neither system shows evidence
for soft X-ray or dust absorption that could explain the differences.

While we measure the source sizes by modeling the data in detail, 
the essence of the measurement is comparing the amplitude of the 
X-ray microlensing variability to that in the optical.  
If the amplitude for the X-ray microlensing
is larger than in the optical, then the X-ray source must be more compact, and
this is what we observed in both \rxj\ and \he.  Given the estimated black
hole mass of $M_{BH} = 2.4 \times 10^9 M_\odot$ (Peng et al. 2006), 
our fits to the X-ray and optical light curves using the microlensing model 
described in \S 2.5 indicate that the 
X-ray emission in \he\ comes from a region smaller than $\sim 6 r_g$
and the UV emission region has a size of  $\sim 30 r_g$. 
We note that the source sizes of \he\ were 
derived using a Bayesian analysis which requires a prior.
As shown in Figures 8 and 9 our upper limit on the X-ray source size 
is robust to prior selection but not the lower limit.
Obtaining a more precise X-ray size estimate requires a more extensive X-ray light curve for the
system.  This is somewhat challenging for \he\ because the long time delay of \he\ 
(162 days, Morgan et al. 2008a) and the intrinsic variability we observe here
means that this lens requires observations separated
by the time delay in order to correct for intrinsic variability.  This is
less of a problem for almost all other candidates for X-ray monitoring because
they have far shorter time delays.  For example, in \rxj\ we observe some 
intrinsic variability but it is significant only on time scales much longer
than the delays between the A/B/C images.  Given the Morgan et al. (2008b)
estimates for PG~1115+080, the Dai et al. (2009) results for \rxj, and our
present results for \he, the X-ray microlensing results suggest that the size
of the X-ray emission arises from close to the inner edge of the 
disk. Thus, our estimated sizes for the X-ray and UV emission regions place significant constraints on AGN models that assume extended
coronal envelopes around accretion disks.
Reducing the uncertainties in these measurements and measuring the scaling of the size
with the estimated masses of the black holes requires longer X-ray time
series and more systems, but there is clearly no technical barrier to making
the measurements and extending them to the dependence of the size on X-ray energy. 

Combining the data from the time series also allows us to search for extended
emission from groups or clusters near the lens, which can be important for
lens models (e.g. Williams et al. 2006, Fassnacht et al. 2007).  Here we
analyzed the X-ray data for the foreground cluster found near \rxj\ by 
Morgan et al. (2006) in more detail, and set limits on the presence of any
other clusters near both systems (see \S 2.4).  
We also used the combined images to search for 
central or odd images, although this is challenging given the resolution
of {\it Chandra} and the expected faintness of the images.  While we
reached 1$\sigma$ limits on the flux ratios of $>100$, significant 
constraints on the central images require limits better by another
order of magnitude (Keeton 2003).  

\acknowledgments
We acknowledge financial support from NASA via the Smithsonian Institution grant SAO SV4-74018. 
CSK acknowledges financial support from NSF grant AST-0708082 and Chandra grants G06-7093 and
GO7-8104.

\small

\clearpage

\normalsize

\beginrefer

\refer Agol, E., \& Krolik, J.\ 1999, ApJ, 524, 49 \\

\refer Anguita, T., Schmidt, R.W., Turner, E.L., Wambsganss, J., Webster, R.L., Loomis, K.A.,
       Long, D., \& McMillan, R., 2008, A\&A, 480, 327 \\

\refer Arnaud, K.~A.\ 1996, ASP 
Conf.~Ser.~101: Astronomical Data Analysis Software and Systems V, 5, 17 \\

\refer Blackburne, J.~A., Pooley, D., \& Rappaport, S.\ 2006, ApJ, 640, 569 \\

\refer Blandford, R.~D., \& McKee, C.~F.\ 1982, ApJ, 255, 419 \\

\refer Chartas, G., et al.\  2000, \apj, 542, 655 \\

\refer Chartas, G., Agol, E., Eracleous, M., Garmire, G., Bautz, M.~W., \& Morgan, N.~D.\ 2002a, ApJ, 68, 509 \\

\refer Chartas, G., Gupta, V., Garmire, G., Jones, C., Falco, E.~E., Shapiro, I.~I., \& Tavecchio, F.\ 2002b, \apj, 565, 96 \\

\refer Chartas, G., Dai, X., Gallagher, S.~C., Garmire, G.~P., Bautz, M.~W., Schechter, P.~L., \& Morgan, N.~D.\ 2001, ApJ, 558, 119 \\

\refer Chartas, G., Eracleous, M., Agol, E., \& Gallagher, S.~C.\ 2004, ApJ, 606, 78 \\

\refer Chartas, G., Eracleous, M., Dai, X., Agol, E., \& Gallagher, S.\ 2007, ApJ, 661, 678 \\

\refer Chartas, G., Falco, E., Forman, W., Jones, C., Schild, R., \& Shapiro, I.\ 1995, ApJ, 445, 140 \\

\refer Dai, X., Kochanek, C.S., Chartas, G., \& Mathur, S., 2006, ApJ, 637, 53 \\

\refer Dai, X., \& Kochanek, C.S., 2008, ApJ submitted [arXiv:0803.1679] \\

\refer Dalal, N., \& Kochanek, C.~S.\ 2002, \apj, 572, 25 \\

\refer Dickey, J.~M., \& Lockman, F.~J.\ 1990, \araa, 28, 215 \\

\refer Fassnacht, C.D., Kocevski, D.D., Auger, M.W., Lubin, L.M., Neureuther, J.L., 
   Jeltema, T.E., Mulchaey, J.S., \& McKean, J.P., 2007, ApJ submitted [arXiv:0711.2066] \\

\refer Garmire, G.~P., Bautz, M.~W., Ford, P.~G., Nousek, J.~A., \& Ricker, G.~R.\ 2003, \procspie, 4851, 28 \\

\refer George, I.~M., \& Fabian, A.~C.\ 1991, \mnras, 249, 352 \\

\refer Gould, A., \& Gaudi, B.~S.\ 1997, ApJ, 486, 687 \\

\refer Gould, A., 2000, ApJ, 535, 928 \\

\refer Grieger, B., Kayser, R., \& Refsdal, S.\ 1988, A\&A, 194, 54 \\

\refer Grieger, B., Kayser, R., \& Schramm, T.\ 1991, A\&A, 252, 508 \\

\refer Haardt, F., \& Maraschi, L., 1991, ApJ, 380, 51 \\

\refer Hirose, S., Krolik, J., De Villiers, J.-P., \& Hawley, J., 2004, ApJ, 606, 1083 \\

\refer Jones, C., \& Forman, W.\ 1999, \apj, 511, 65 \\

\refer Jovanovi{\'c}, P., Zakharov, A.~F., Popovi{\'c}, L.~{\v C}., 
\& Petrovi{\'c}, T.\ 2008, \mnras, 386, 397 \\

\refer Keeton, C.R., 2003 ApJ, 582 17 \\

\refer Kochanek, C.~S., Dai, X., Morgan, C., Morgan, N., Poindexter, S.~C.~G. 
\& Chartas, G., \ 2007, Statistical Challenges in Modern Astronomy IV, 371, 43 \\

\refer Kochanek, C.S., Morgan, N.D., Falco, E.E., et al., 2006, ApJ, 640, 47 \\

\refer Kochanek, C.~S.\ 2002, \apj, 578, 25 \\

\refer Kochanek, C.~S.\ 2004, \apj, 605, 58 \\

\refer Leh{\'a}r, J., et al.\ 2000, \apj, 536, 584 \\

\refer Lewis, G.~F., Irwin, 
M.~J., Hewett, P.~C., \& Foltz, C.~B.\ 1998, \mnras, 295, 573  \\

\refer Machida, M., \& Matsumoto, R., 2003, ApJ, 585, 429 \\

\refer Mao, S., \& Schneider, P.\ 1998, \mnras, 295, 587 \\

\refer Merloni, A., 2003, MNRAS, 341, 1051 \\

\refer Mineshige, S., \& Yonehara, A.\ 1999, PASJ, 51, 497 \\

\refer Morgan, N.~D., Chartas, G., Malm, M., Bautz, M.~W., Burud, I., Hjorth, J., Jones, S.~E., \& Schechter, P.~L.\ 2001, ApJ, 555, 1 \\

\refer Morgan, N.~D., Kochanek, C.~S., Falco, E.~E., \& Dai, X.\ 2006, ArXiv Astrophysics e-prints, arXiv:astro-ph/0605321 \\

\refer Morgan, C.~W., Eyler, M.~E., Kochanek, C.~S., Morgan, N.~D., Falco, E.~E., Vuissoz, C., Courbin, F., \& Meylan, G., 2008, ApJ, 676, 80 \\


\refer Morgan, C.~W., Eyler, 
M.~E., Kochanek, C.~S., Morgan, N.~D., Falco, E.~E., Vuissoz, C., Courbin, 
F., \& Meylan, G.\ 2008a, \apj, 676, 80 \\

\refer Morgan, C.~W., Kochanek, 
C.~S., Morgan, N.~D., 
\& Falco, E.~E.\ 2007, ArXiv e-prints, 707, arXiv:0707.0305 \\

\refer Morgan, C.~W., Kochanek, C.~S., Dai, X., Morgan, N.~D., \& Falco, E.~E.\ 2008b, ArXiv e-prints, 802, arXiv:0802.1210 \\

\refer Mori, K., Tsunemi, H., Miyata, E., Baluta, C., Burrows, D. N.,
Garmire, G. P., \& Chartas, G. 2001, in ASP Conf. Ser. 251, New Century
of X-Ray Astronomy, ed. H. Inoue \& H. Kunieda (San Francisco: ASP), 576 \\

\refer Mortonson, M.J., Schechter, P.L., \& Wambsganss, J., 2005, ApJ, 628, 594 \\

\refer Mushotzky, R.~F., Done, C., \& Pounds, K.~A.\ 1993, \araa, 31, 717 \\

\refer Netzer, H., \& Peterson, B.~M.\ 1997, Astronomical Time Series, 218, 85 \\

\refer Ofek, E.O. \& Maoz, D., 2003, ApJ, 594, 101 \\

\refer Peng, C.Y., Impey, C.D., Rix, H.-W., Kochanek, C.S., Keeton, C.R., Falco, E.E., L\'ehar, J., \& Mcleod, B.A., 2006, ApJ 649, 616 \\

\refer Peterson, B.~M.\ 1993, PASP, 105, 247 \\

\refer Poindexter, S., Morgan, N., Kochanek, C.S., \& Falco, E.E., 2007, ApJ, 660, 146 \\

\refer Pointdexter, S. Mogan, N., \& Kochanek, C.S., 2008, ApJ, 673, 34 \\

\refer Pooley, D., Blackburne, J.~A., Rappaport, S., Schechter, P.~L., \& Fong, W.-f.\ 2006, ApJ, 648, 67 \\

\refer Pooley, D., Blackburne,  J.~A., Rappaport, S., \& Schechter, P.~L.\ 2007, ApJ, 661, 19 \\

\refer Popovi{\'c}, L.~{\v C}., \& Chartas, G.\ 2005, MNRAS, 357, 135 \\

\refer Popovi{\'c}, L.~{\v C}., Mediavilla, E.~G., Jovanovi{\'c}, P., \& Mu{\~n}oz, J.~A.\ 2003, A\&A, 398, 975 \\

\refer Reynolds, C.S. \& Nowak, M.A., 2003, PhR, 377, 389\\

\refer Risaliti, G., Elvis,  M., Fabbiano, G., Baldi, A., Zezas, A., \& Salvati, M.\ 2007, ApJL, 659, L111 \\

\refer Schechter, P.~L., \& Wambsganss, J.\ 2002, \apj, 580, 685 \\

\refer Schneider, P., Ehlers, J., \& Falco, E.~E.\ 1992, Gravitational Lenses, XIV,~Springer-Verlag Berlin Heidelberg New York.~ Also Astronomy and Astrophysics Library. \\

\refer Shankar, F., Weinberg, 
D.~H., \& Miralda-Escude', J.\ 2007, ArXiv e-prints, 710, arXiv:0710.4488 \\

\refer Tsunemi, H., Mori, K., 
Miyata, E., Baluta, C., Burrows, D.~N., Garmire, G.~P., \& Chartas, G.\ 
2001, \apj, 554, 496 \\

\refer Williams, K.A., Momchemva, I., Keeton, C.R., Zabludoff, AI., \& L\'ehar, J., 2006, ApJ, 646, 85

\refer Wise, M. W., Davis, J. E., Huenemoerder, Houck, J. C., Dewey, D.
Flanagan, K. A., and Baluta, C. 1997,
{\it The MARX 3.0 User Guide, CXC Internal Document}
available at http://space.mit.edu/ASC/MARX/ \\

\refer Wryzkwoski, L., et al., 2003 Acta Astron. 53 229 \\

\refer Yonehara, A., Mineshige, S., Fukue, J., Umemura, M., \& Turner, E.~L.\ 1999, A\&A, 343, 41 \\

\refer Young, A.~J., \& Reynolds, C.~S.\ 2000, ApJ, 529, 101 \\

\endrefer

\clearpage
\scriptsize
\begin{center}
\begin{tabular}{clcccccc}
\multicolumn{8}{c}{TABLE 1}\\
\multicolumn{8}{c}{Log of Observations of Quasars \rxj\ and \he} \\
 & & && & & &\\ \hline\hline
         &         & {\it Chandra}                                      & Exposure & & & &      \\
Epoch & Observation  &  Observation  &   Time   & $N_{\rm A}$$\tablenotemark{a}$ & $N_{\rm B}$$\tablenotemark{a}$ & $N_{\rm C}$$\tablenotemark{a}$ & $N_{\rm D}$ $\tablenotemark{a}$   \\
& Date           &                       ID                &   (ks)       & counts   & counts & counts & counts   \\
&   &    &    &   &  &  & \\
\multicolumn{8}{c}{Observations of \rxj}\\
\hline
1&2004 April 12     &  4814           & 10.  &       425 $\pm$ 22 & 2950 $\pm$ 54 &  839 $\pm$ 29 & 211 $\pm$ 15 \\
2&2006 March 10     &  6913           & 4.9   &       393 $\pm$ 20 & 624 $\pm$ 25 & 204 $\pm$ 14 &  103 $\pm$ 10   \\
3&2006 March 15   &  6912            &  4.4  &       381 $\pm$ 20 & 616 $\pm$ 25 & 233 $\pm$ 15 & 93 $\pm$  10 \\
4&2006 April 12         &  6914           &  4.9   &     413  $\pm$ 20 & 507 $\pm$ 23 & 146 $\pm$ 12  & 131 $\pm$ 12 \\
5&2006 November 10         &  6915           &  4.8   &     3708  $\pm$ 61 & 1411 $\pm$ 38 & 367 $\pm$ 19  & 155 $\pm$ 13 \\
6&2006 November 13       &  6916           &  4.8   &     3833  $\pm$ 62 & 1618 $\pm$ 40 & 415 $\pm$ 20  & 115 $\pm$ 11 \\
 & && & & & &\\ \hline \\
\multicolumn{8}{c}{Observations of \he} \\
\hline
1&2000 June 10                 & 375    & 47.4 & 1024 $\pm$ 32   & 547 $\pm$ 23   \\
2&2006 February 16          &  6918 & 5.0   &  38 $\pm$  6&  86 $\pm$ 9 \\
3&2006 March 15              & 6917 &  4.6  &  40 $\pm$ 6 & 59 $\pm$ 8\\
4&2006 April 09               & 6919  &  4.9  &  33 $\pm$ 6 &  61 $\pm$ 8 \\
5&2006 October 31           & 6920  &   5.0  & 60 $\pm$ 8 & 40 $\pm$ 6\\
6&2006 November 08       & 6921  &   4.9  & 52 $\pm$ 7 & 57 $\pm$ 8 \\

\hline \hline
\end{tabular}
\end{center}
${}^{a}${Background-subtracted source counts for events with energies in the 0.2--10~keV band.
The counts for images B and C for the 2004 April 12 observation of \rxj\ are corrected for pile-up by factors of 37\% and 16\%, respectively.
The counts for images A and B  for the 2006 November 10 observation of \rxj\ are corrected for pile-up by factor of 32\% and 13\%, respectively.
The counts for images A and B for the 2006 November 13 observation of \rxj\ are corrected for pile-up by factors of 33\% and 17\%, respectively. } \\

\clearpage
\scriptsize
\begin{center}
\begin{tabular}{ccccc}
\multicolumn{5}{c}{TABLE 2}\\
\multicolumn{5}{c}{OPTICAL AND X-RAY OFFSETS OF \rxj\ AND \he\ IMAGES} \\
& & & & \\ \hline\hline
\multicolumn{1}{c} {Telescope} &
\multicolumn{1}{c} {C} &
\multicolumn{1}{c} {A} &
\multicolumn{1}{c} {B} &
\multicolumn{1}{c} {D} \\
\multicolumn{1}{c} {} &
\multicolumn{1}{c} {$\Delta\alpha$($''$), $\Delta\delta$($''$)} &
\multicolumn{1}{c} {$\Delta\alpha$($''$), $\Delta\delta$($''$)} &
\multicolumn{1}{c} {$\Delta\alpha$($''$), $\Delta\delta$($''$)} &
\multicolumn{1}{c} {$\Delta\alpha$($''$), $\Delta\delta$($''$)} \\ \hline
& & & & \\
\multicolumn{5}{c}{\rxj\ Image Offsets}\\
\hline
HST     &0,0   &0.588$\pm$0.003,1.120$\pm$0.003 &0.618$\pm$0.003,2.307$\pm$0.003 &$-$2.517$\pm$0.003, 1.998$\pm$0.003    \\
{\it Chandra} &0,0   &0.59$\pm$0.02, 1.12$\pm$0.02 & 0.62$\pm$0.02,  2.31$\pm$0.02    &$-$2.52$\pm$0.02, 2.00$\pm$0.02    \\
\hline
& & & & \\
\multicolumn{5}{c}{\he\ Image Offsets}\\
\hline
HST     & ---   &0,0 &2.901$\pm$0.003,$-$1.332$\pm$0.003 & ---  \\
{\it Chandra} & ---  & 0,0 & 2.92$\pm$0.02,  $-$1.33$\pm$0.02    & ---    \\
\hline \hline
\end{tabular}
\end{center}

\clearpage
\scriptsize
\begin{center}
\begin{tabular}{cccccc}
\multicolumn{6}{c}{TABLE 3}\\
\multicolumn{6}{c}{RESULTS FROM FITS TO THE {\it Chandra} IMAGE SPECTRA OF \rxj} \\
 & & & & & \\ \hline\hline
\multicolumn{1}{c} {Epoch} &
\multicolumn{1}{c} {Parameter$^{a}$} &
\multicolumn{1}{c} {Values For} &
\multicolumn{1}{c} {Values For} &
\multicolumn{1}{c} {Values For} &
\multicolumn{1}{c} {Values For} \\

        &           & Image A$^{b}$  & Image B$^{b}$&                             Image C$^{b}$                 &       Image D$^{b}$      \\
    &           &   &&                                              &       \\                        
 1 &$\Gamma$     &  1.44$_{-0.08}^{+0.08}$ &    1.41$_{-0.07}^{+0.07}$  &  1.47$_{-0.12}^{+0.12}$      &   1.95$_{-0.20}^{+0.21}$ \\ 
 & 0.2--2~keV Flux (10$^{-13}$~erg~s$^{-1}$~cm$^{-2}$) & 1.4$_{-0.2}^{+0.2}$ & 9.1$_{-0.8}^{+0.8}$  & 3.1$_{-0.5}^{+0.5}$ &0.9$_{-0.2}^{+0.2}$ \\
  & 2--10~keV Flux (10$^{-13}$~erg~s$^{-1}$~cm$^{-2}$) & 2.7$_{-0.4}^{+0.4}$ & 15.7$_{-1.4}^{+1.4}$ & 3.8$_{-0.5}^{+0.6}$& 0.7$_{-0.2}^{+0.2}$ \\
 &$C-statistic/nbins$ & 556/787                      &   661/787                      &    583/787                          &   368/787                            \\
  &$\chi^2/{\nu}$ & 25/31                              &   105/97                       &  28/41                               &   15.4/15                               \\
&$P(\chi^2/{\nu})$$^{c}$& 0.76                    &  0.27                           &     0.95                             &0.42                                 \\  
     &           &   &&                                              &      \\                        
 2 &$\Gamma$     &  1.58$_{-0.15}^{+0.15}$ & 1.63$_{-0.11}^{+0.11}$ &   1.62$_{-0.19}^{+0.19}$     &  1.66$_{-0.27}^{+0.28}$  \\ 
 & 0.2--2~keV Flux (10$^{-13}$~erg~s$^{-1}$~cm$^{-2}$) & 2.9$_{-0.6}^{+0.6}$ & 5.0$_{-0.8}^{+0.8}$  & 1.3$_{-0.3}^{+0.3}$ &0.8$_{-0.4}^{+0.4}$ \\
  & 2--10~keV Flux (10$^{-13}$~erg~s$^{-1}$~cm$^{-2}$) & 4.5$_{-0.9}^{+1.2}$ & 7.0$_{-1.1}^{+1.2}$ & 2.3$_{-0.6}^{+0.7}$& 1.1$_{-0.4}^{+0.5}$ \\
 &$C-statistic/nbins$ & 531/838                       & 653/838                         & 435/838                             & 300/838                              \\
  &$\chi^2/{\nu}$ & 33/28                              & 61/52                            & 14.4/16                                & 2.4/6                                 \\
&$P(\chi^2/{\nu})$$^{c}$& 0.24                    &0.18                                 & 0.57                                    &0.9                                   \\
    &           &   &&                                              &       \\                        
 3 &$\Gamma$     &  1.61$_{-0.15}^{+0.15}$ & 1.73$_{-0.11}^{+0.11}$ &   1.57$_{-0.18}^{+0.18}$     &  1.60$_{-0.28}^{+0.28}$ \\ 
 & 0.2--2~keV Flux (10$^{-13}$~erg~s$^{-1}$~cm$^{-2}$) & 3.1$_{-0.6}^{+0.8}$ & 5.6$_{-1.5}^{+1.5}$  & 1.8$_{-0.5}^{+0.5}$ &0.8$_{-0.3}^{+0.4}$ \\
  & 2--10~keV Flux (10$^{-13}$~erg~s$^{-1}$~cm$^{-2}$) & 4.6 $_{-0.9}^{+1.2}$& 6.7$_{-1.2}^{+1.3}$ & 3.0$_{-0.8}^{+0.9}$& 1.2$_{-0.5}^{+0.6}$ \\
 &$C-statistic/nbins$ & 521/838                       & 555/838                         & 425/838                             & 292/838                              \\
  &$\chi^2/{\nu}$ & 26/27                              & 42/48                            & 18.5/17                                & 3.8/6                                 \\
&$P(\chi^2/{\nu})$$^{c}$& 0.52                    &0.73                                 & 0.36                                    &0.7                                   \\
    &           &   &&                                              &       \\                        
 4 &$\Gamma$     &  1.53$_{-0.14}^{+0.14}$ & 1.64$_{-0.12}^{+0.12}$ &   1.70$_{-0.23}^{+0.23}$     &  1.62$_{-0.24}^{+0.24}$  \\ 
 & 0.2--2~keV Flux (10$^{-13}$~erg~s$^{-1}$~cm$^{-2}$) & 3.0$_{-0.6}^{+0.6}$ & 3.9$_{-0.7}^{+0.7}$  & 1.2$_{-0.4}^{+0.4}$ &1.0$_{-0.4}^{+0.4}$ \\
  & 2--10~keV Flux (10$^{-13}$~erg~s$^{-1}$~cm$^{-2}$) & 5.1$_{-1.0}^{+1.1}$ & 5.4$_{-1.0}^{+1.2}$ & 1.5$_{-0.5}^{+0.6}$ & 1.5$_{-0.5}^{+0.6}$ \\
 &$C-statistic/nbins$ & 541/838                       & 548/838                         & 333/838                             & 350/838                              \\
  &$\chi^2/{\nu}$ & 27/29                              & 40/41                            & 7.7/10                                & 13/9                                \\
&$P(\chi^2/{\nu})$$^{c}$& 0.58                    &0.52                                 & 0.66                                    &0.15                                  \\
    &           &   &&                                              &       \\    
     5 &$\Gamma$     &  1.59$_{-0.07}^{+0.07}$ & 1.82$_{-0.09}^{+0.08}$ &   1.90$_{-0.15}^{+0.15}$     &  1.54$_{-0.22}^{+0.22}$  \\ 
 & 0.2--2~keV Flux (10$^{-13}$~erg~s$^{-1}$~cm$^{-2}$) & 29$_{-2}^{+2}$ & 13$_{-1}^{+1}$  & 3.4$_{-0.5}^{+0.5}$ &1.1$_{-0.3}^{+0.3}$ \\
  & 2--10~keV Flux (10$^{-13}$~erg~s$^{-1}$~cm$^{-2}$) & 35$_{-4}^{+3}$ & 11$_{-1}^{+2}$ & 2.8$_{-0.6}^{+0.7}$ & 1.9$_{-0.5}^{+0.6}$ \\
 &$C-statistic/nbins$ & 771/838                       & 626/838                         & 502/838                             & 402/838                             \\
  &$\chi^2/{\nu}$ & 72/59                              & 66/63                       & 34/28                                & 7/11                                \\
&$P(\chi^2/{\nu})$$^{c}$& 0.1                    &0.37                                 & 0.19                                    &0.7                                  \\
    &           &   &&                                              &       \\                        
 6 &$\Gamma$     &  1.51$_{-0.07}^{+0.07}$ & 1.82$_{-0.08}^{+0.08}$ &   1.76$_{-0.14}^{+0.14}$     &  2.28$_{-0.28}^{+0.29}$  \\ 
 & 0.2--2~keV Flux (10$^{-13}$~erg~s$^{-1}$~cm$^{-2}$) & 30$_{-2}^{+2}$ & 15$_{-2}^{+2}$  & 3.2$_{-0.5}^{+0.5}$ &1.4$_{-0.5}^{+0.5}$ \\
  & 2--10~keV Flux (10$^{-13}$~erg~s$^{-1}$~cm$^{-2}$) & 36$_{-3}^{+3}$ & 13$_{-2}^{+2}$ & 3.6$_{-0.8}^{+0.9}$ & 0.6$_{-0.2}^{+0.3}$ \\
 &$C-statistic/nbins$ & 768/838                       & 658/838                         & 489/838                             & 276/838                            \\
  &$\chi^2/{\nu}$ & 65/59                              & 62/70                           & 29/33                                & 9/8                                \\
&$P(\chi^2/{\nu})$$^{c}$& 0.29                    &0.76                                & 0.66                                    &0.36                                \\
    &           &   &&                                              &       \\                        
                    
\hline \hline
\end{tabular}
\end{center}
\noindent
${}^{a}$ Model 1 consists of a power-law modified by Galactic absorption. \\
${}^{b}$All errors are for 90\% confidence unless mentioned otherwise with all
parameters taken to be of interest except absolute normalization.\\
${}^{c}$$P(\chi^2/{\nu})$ is the probability of exceeding $\chi^{2}$ for ${\nu}$ degrees of freedom
if the model is correct.\\

\clearpage
\scriptsize
\begin{center}
\begin{tabular}{cccc}
\multicolumn{4}{c}{TABLE 4}\\
\multicolumn{4}{c}{RESULTS FROM FITS TO THE {\it Chandra} IMAGE SPECTRA OF \he} \\
 & & & \\ \hline\hline
\multicolumn{1}{c} {Epoch} &
\multicolumn{1}{c} {Parameter$^{a}$} &
\multicolumn{1}{c} {Values For} &
\multicolumn{1}{c} {Values For} \\

        &           & Image A$^{b}$  & Image B$^{b}$            \\
    &           &   &                                          \\   
1 &$\Gamma$     &  1.59$_{-0.08}^{+0.08}$ &   1.86$_{-0.12}^{+0.12}$  \\ 
 & 0.2--2~keV Flux (10$^{-14}$~erg~s$^{-1}$~cm$^{-2}$) &  6.9$_{-0.6}^{+0.6}$ & 4.28$_{-0.6}^{+0.6}$  \\
  & 2--10~keV Flux (10$^{-14}$~erg~s$^{-1}$~cm$^{-2}$) & 10.6$_{-0.1}^{+0.2}$ & 3.97$_{-0.8}^{+0.7}$\\
 &$C-statistic/nbins$ &       623/790                 &              \\
     &           &   &                                                \\                           
 2 &$\Gamma$     &  1.4$_{-0.4}^{+0.4}$ &   1.6$_{-0.3}^{+0.3}$  \\ 
 & 0.2--2~keV Flux (10$^{-14}$~erg~s$^{-1}$~cm$^{-2}$) & 2.6$_{-1.6}^{+1.8}$ &6.5$_{-2.7}^{+3.0}$  \\
  & 2--10~keV Flux (10$^{-14}$~erg~s$^{-1}$~cm$^{-2}$) & 5.8$_{-3.5}^{+4.6}$ &10.2$_{-4.0}^{+6.0}$\\
 &$C-statistic/nbins$ & 192/838                      &   284/1060          \\
     &           &   &                                                \\  
3 &$\Gamma$     &  1.53$_{-0.41}^{+0.40}$ &   2.4$_{-0.4}^{+0.4}$ \\ 
 & 0.2--2~keV Flux (10$^{-14}$~erg~s$^{-1}$~cm$^{-2}$) &3.4$_{-2.3}^{+2.2}$&8.4$_{-4.1}^{+3.8}$\\
  & 2--10~keV Flux (10$^{-14}$~erg~s$^{-1}$~cm$^{-2}$) & 5.9$_{-3.0}^{+4.0}$&  3.0$_{-1.5}^{+2.3}$\\
 &$C-statistic/nbins$ & 186/838                       &   195/1060         \\
     &           &   &                                                \\
4 &$\Gamma$     &  1.76$_{-0.47}^{+0.50}$ &   2.03$_{-0.36}^{+0.37}$   \\ 
 & 0.2--2~keV Flux (10$^{-14}$~erg~s$^{-1}$~cm$^{-2}$) &2.9$_{-1.9}^{+2.3}$&  6.5$_{-2.7}^{+3.3}$\\
  & 2--10~keV Flux (10$^{-14}$~erg~s$^{-1}$~cm$^{-2}$) & 3.2$_{-2.0}^{+4.0}$&  4.3$_{-2.1}^{+3.5}$\\
 &$C-statistic/nbins$ & 160/838                       &   224/1060         \\
     &           &   &                                                 \\
5 &$\Gamma$     &  1.59$_{-0.34}^{+0.35}$ &   2.33$_{-0.46}^{+0.49}$  \\ 
 & 0.2--2~keV Flux (10$^{-14}$~erg~s$^{-1}$~cm$^{-2}$) &4.5$_{-2.3}^{+2.5}$& 5.0$_{-2.8}^{+3.5}$ \\
  & 2--10~keV Flux (10$^{-14}$~erg~s$^{-1}$~cm$^{-2}$) & 6.9$_{-3.2}^{+5.0}$ & 1.8$_{-1.1}^{+2.0}$\\
 &$C-statistic/nbins$ & 236/838                     &   165/1060        \\
     &           &   &                                                 \\
6 &$\Gamma$     &  1.61$_{-0.35}^{+0.36}$ &   1.81$_{-0.37}^{+0.42}$  \\ 
 & 0.2--2~keV Flux (10$^{-14}$~erg~s$^{-1}$~cm$^{-2}$) &4.4$_{-2.6}^{+2.6}$& 4.6$_{-2.3}^{+2.4}$ \\
  & 2--10~keV Flux (10$^{-14}$~erg~s$^{-1}$~cm$^{-2}$) & 6.4$_{-3.1}^{+4.7}$& 4.7$_{-2.4}^{+3.6}$ \\
 &$C-statistic/nbins$ & 230/838                       &   222/1060       \\
     &           &     &                                                    \\
\hline \hline
\end{tabular}
\end{center}
\noindent
${}^{a}$ Model 1 consists of a power-law modified by Galactic absorption. The X-ray fluxes represent unabsorbed values. \\
${}^{b}$All errors are for 90\% confidence unless mentioned otherwise with all
parameters taken to be of interest except absolute normalization.\\

\clearpage
\begin{figure*}
\centerline{\includegraphics[width=9.5cm]{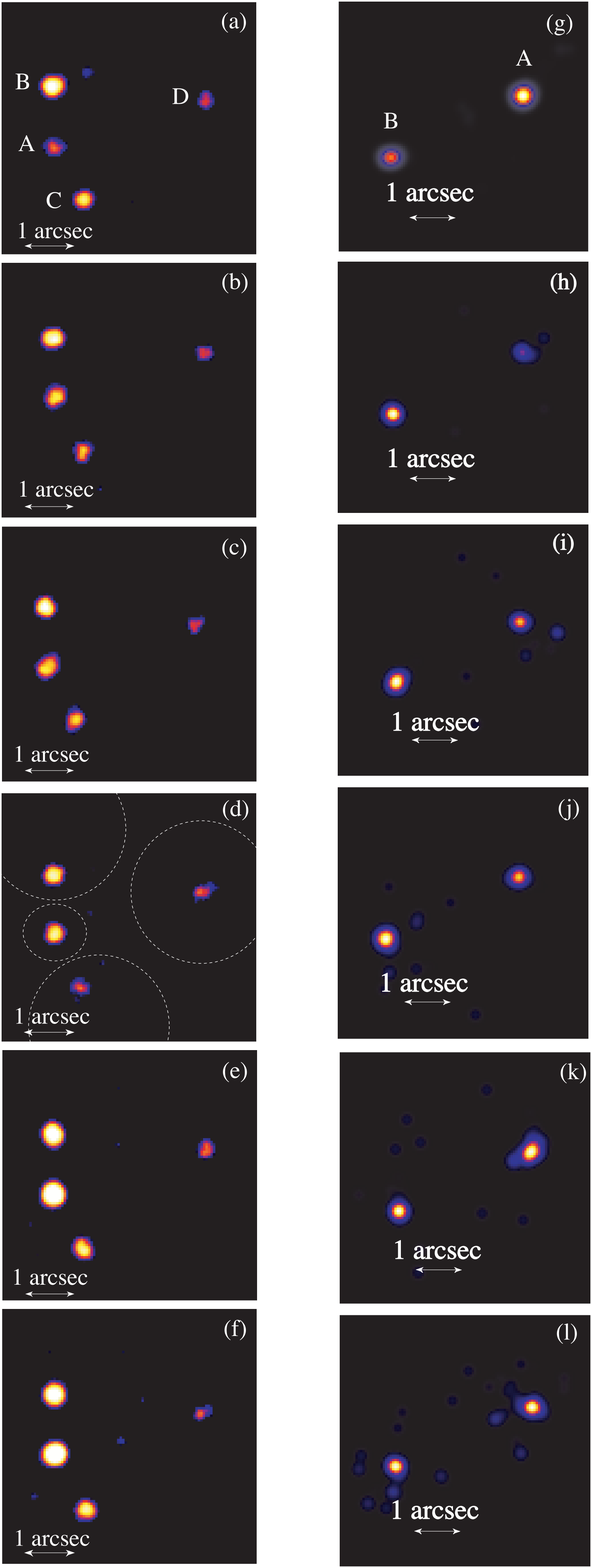}}
\caption{Lucy-Richardson deconvolved images in the 0.2 - 10 keV bandpass for the \chandra\ observations of \rxj\ (left) and \he\  (right). The brightness scale of the images is proportional to the count rate.  
The images are displayed with a linear brightness scale. East is to the left and North is up.
 \label{f1.eps}}
\end{figure*}

\begin{figure*}
\centerline{\includegraphics[width=10cm]{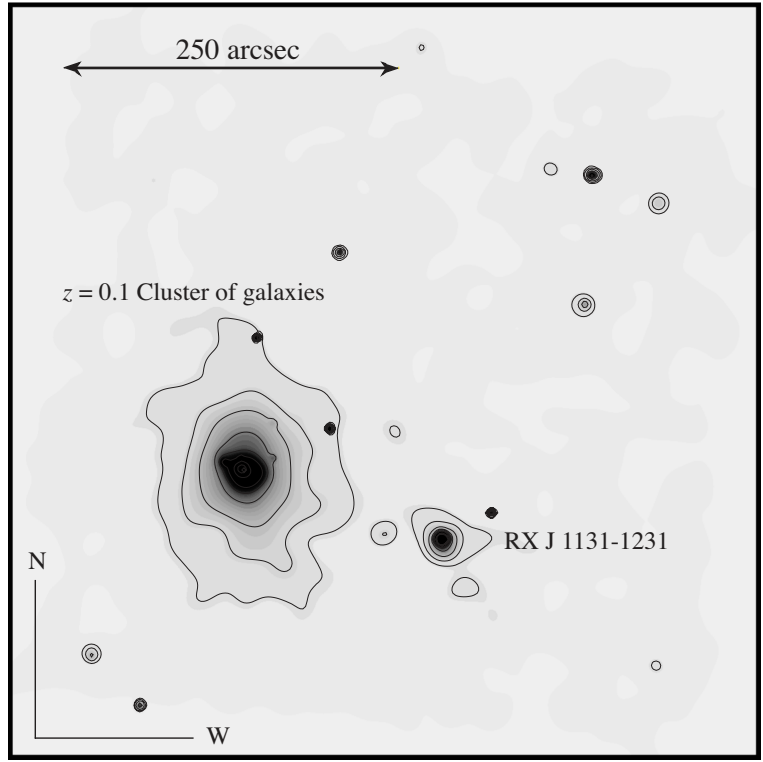}}
\caption{Adaptively smoothed image of the 10~ks 2004 \chandra\ observation of \rxj.
To reduce background contamination and to enhance possible soft extended X-ray emission 
we filtered the image to include only photons with energies ranging between 0.4 and 3.0~keV. 
 \label{f2.eps}}
\end{figure*}

\begin{figure*}
\centerline{\includegraphics[width=10cm]{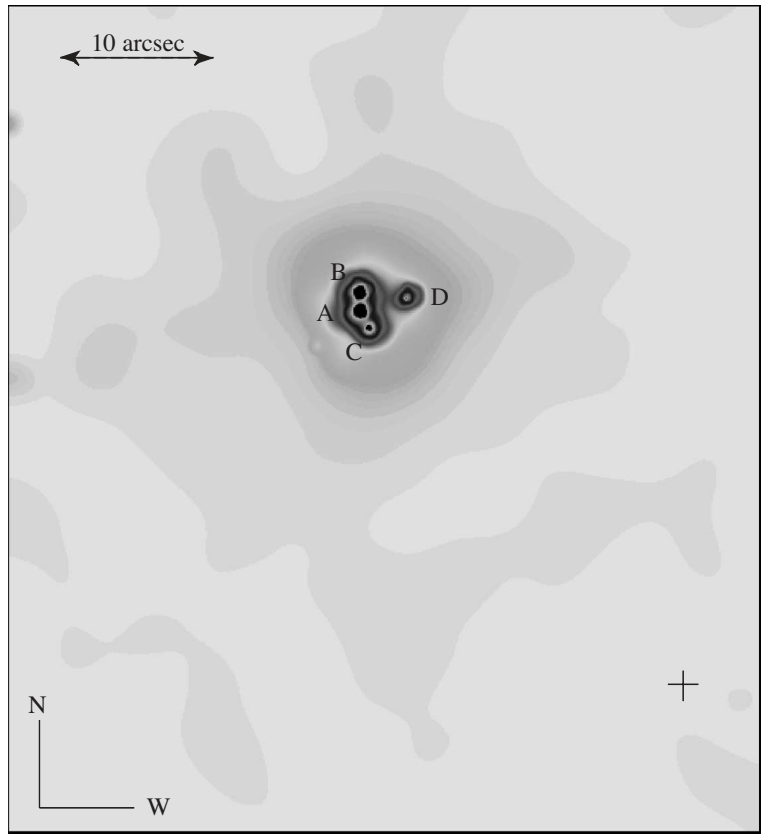}}
\caption{Stacked image of the six observations of \rxj\ listed in Table 1.
The stacked image was filtered to include only photons with energies ranging between 0.4 and 3.0~keV and adaptively smoothed. 
We have indicated with a cross the location of the reported, but low significance, extended X-ray emission claimed to possibly originate from a lensing cluster of galaxies. 
There is no indication of any extended emission.
\label{f3.eps}}
\end{figure*}

\begin{figure*}
\centerline{\includegraphics[width=10cm]{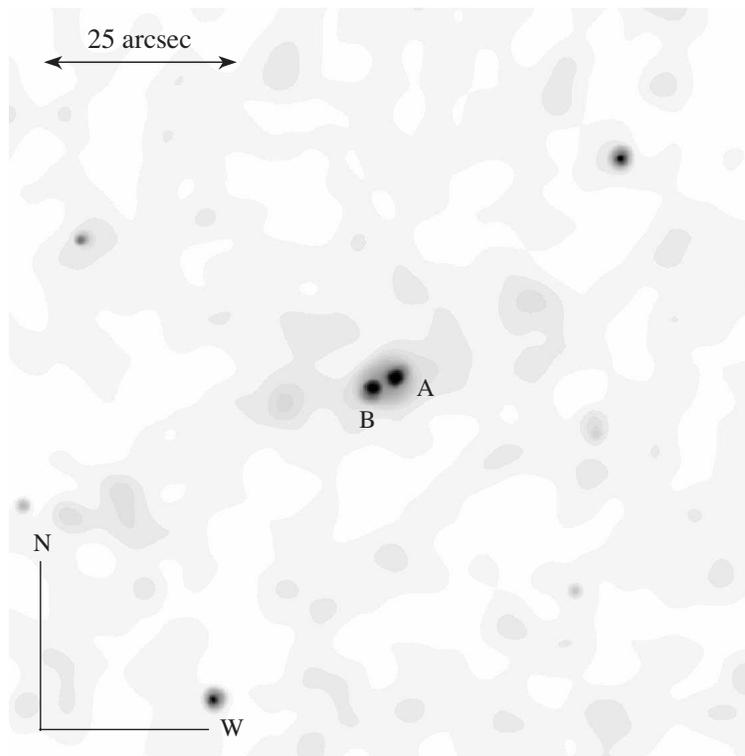}}
\caption{Stacked image of the six observations of \he\ listed in Table 1.
The stacked image was filtered to include only photons with energies ranging between 0.4 and 3.0~keV and adaptively smoothed. 
There is no indication of any extended emission originating from a group or cluster of galaxies.
 \label{f4.eps}}
\end{figure*}

\begin{figure*}
\centerline{\includegraphics[width=10cm]{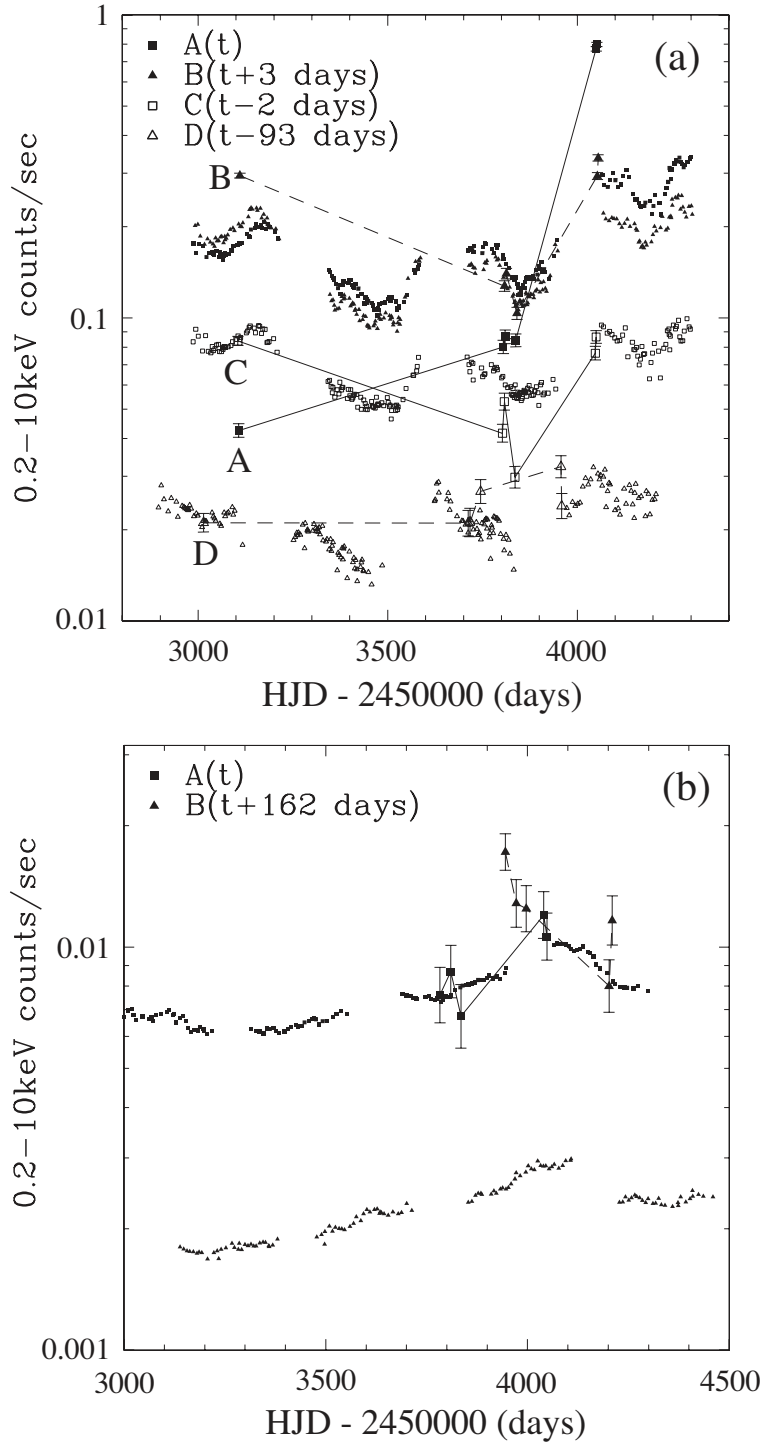}}
\caption{ 
The 0.2--10~keV (large point with error bars) and $R$-band (small points) light curves corrected for the 
time delay of (a) images A, B, C and D  of \rxj\ and 
(b) images A and B of \he, respectively.
The optical fluxes of \rxj\ are normalized such that the 
normalized optical flux of image D is equal to the 0.2--10~keV count-rate of image D near day 3000.
The optical fluxes of \he\ are normalized such that the 
normalized optical flux of image A is equal to the 0.2--10~keV count-rate of image A near day 3780.
 \label{f5.eps}}
\end{figure*}


\begin{figure*}
\centerline{\includegraphics[width=10cm]{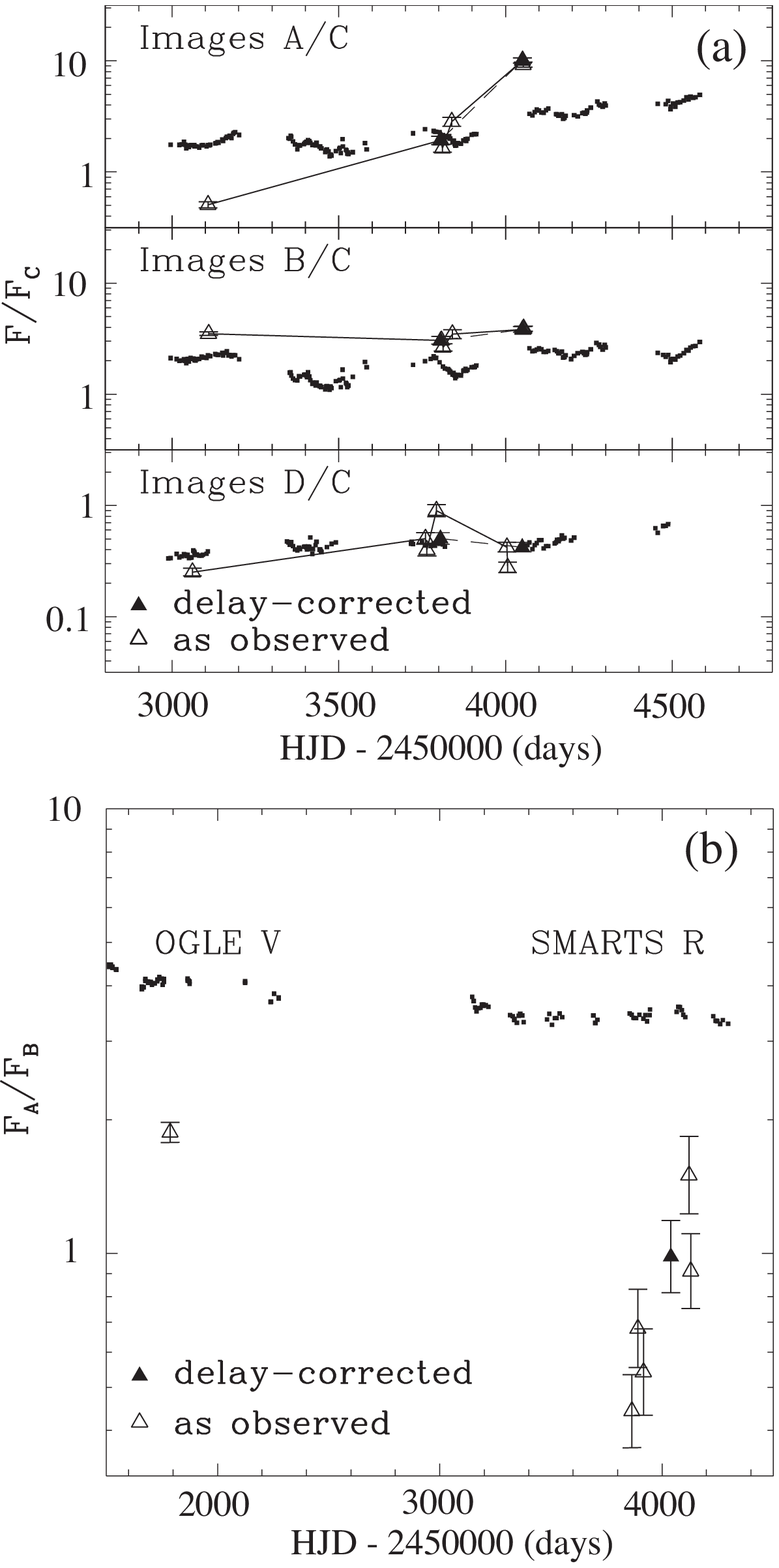}}
\caption{ (a) X-ray (0.2--10 keV, large points) and optical ($R$-band, small points) flux ratios 
F$_{\rm A}$/F$_{\rm C}$, F$_{\rm B}$/F$_{\rm C}$ and F$_{\rm D}$/F$_{\rm C}$
for  \rxj. (b) X-ray (0.2--10 keV) and optical ($R$-band) flux ratios F$_{\rm A}$/F$_{\rm B}$ for  \he.
The optical flux ratios of \rxj\ and \he\ are corrected for the time-delays.
The 0.2--10~keV time-delay corrected flux ratios of \rxj\ and \he\ are shown with solid triangles
and the observed ratios are shown with open triangles. 
 \label{f6.eps}}
\end{figure*}


\begin{figure*}
\centerline{\includegraphics[width=18cm]{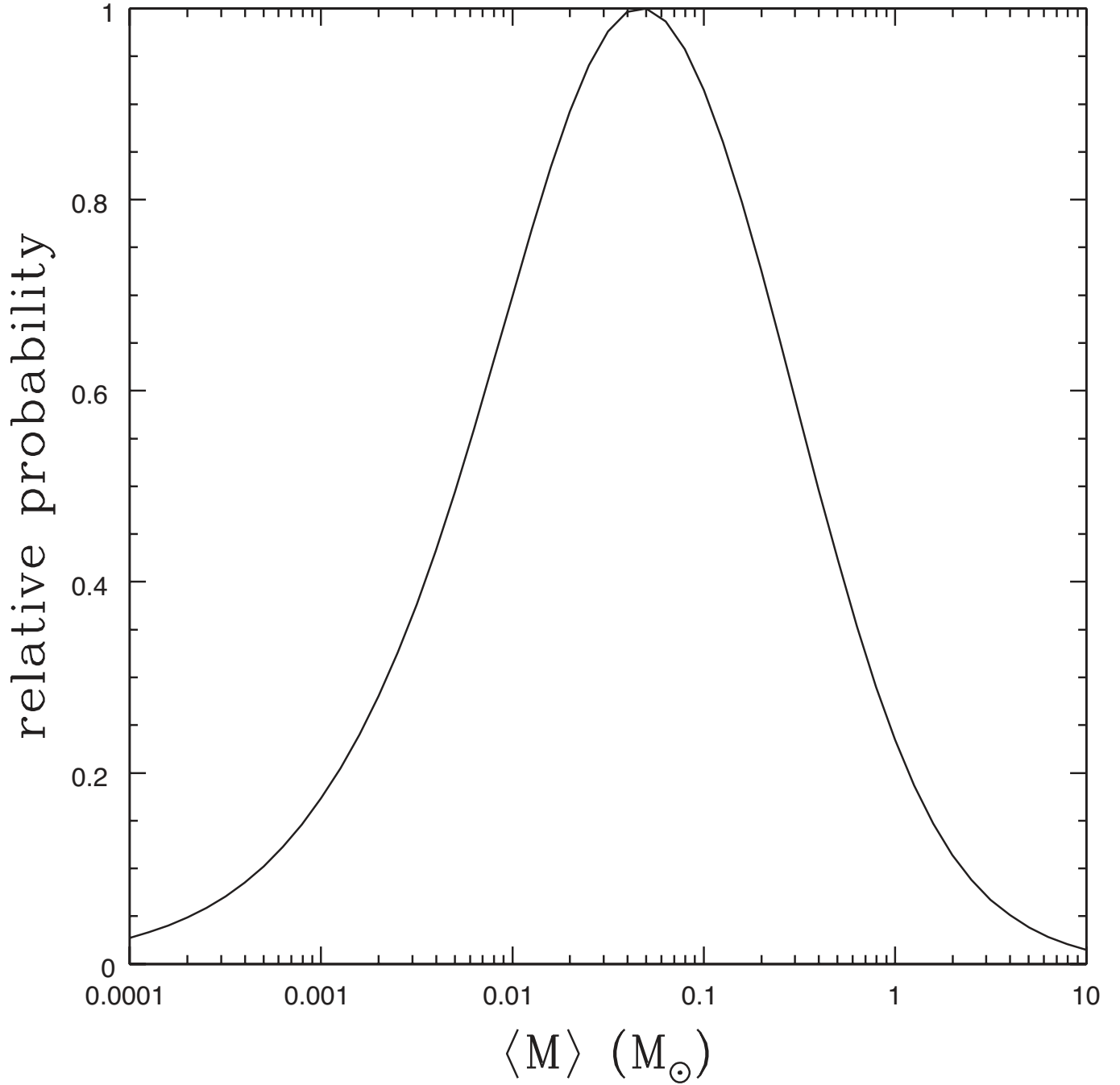}}
\caption{Probability density of the microlens mass scale $\langle M\rangle$ given
our model for the estimated physical velocities.  The peak is consistent with our
alternate prior on the mass scale ($0.1M_\odot < \langle M\rangle < M_\odot$) but
broader because it is difficult to tightly constrain the mass scale using the 
velocities. 
\label{f7.eps}}
\end{figure*}

\begin{figure*}
\centerline{\includegraphics[width=18cm]{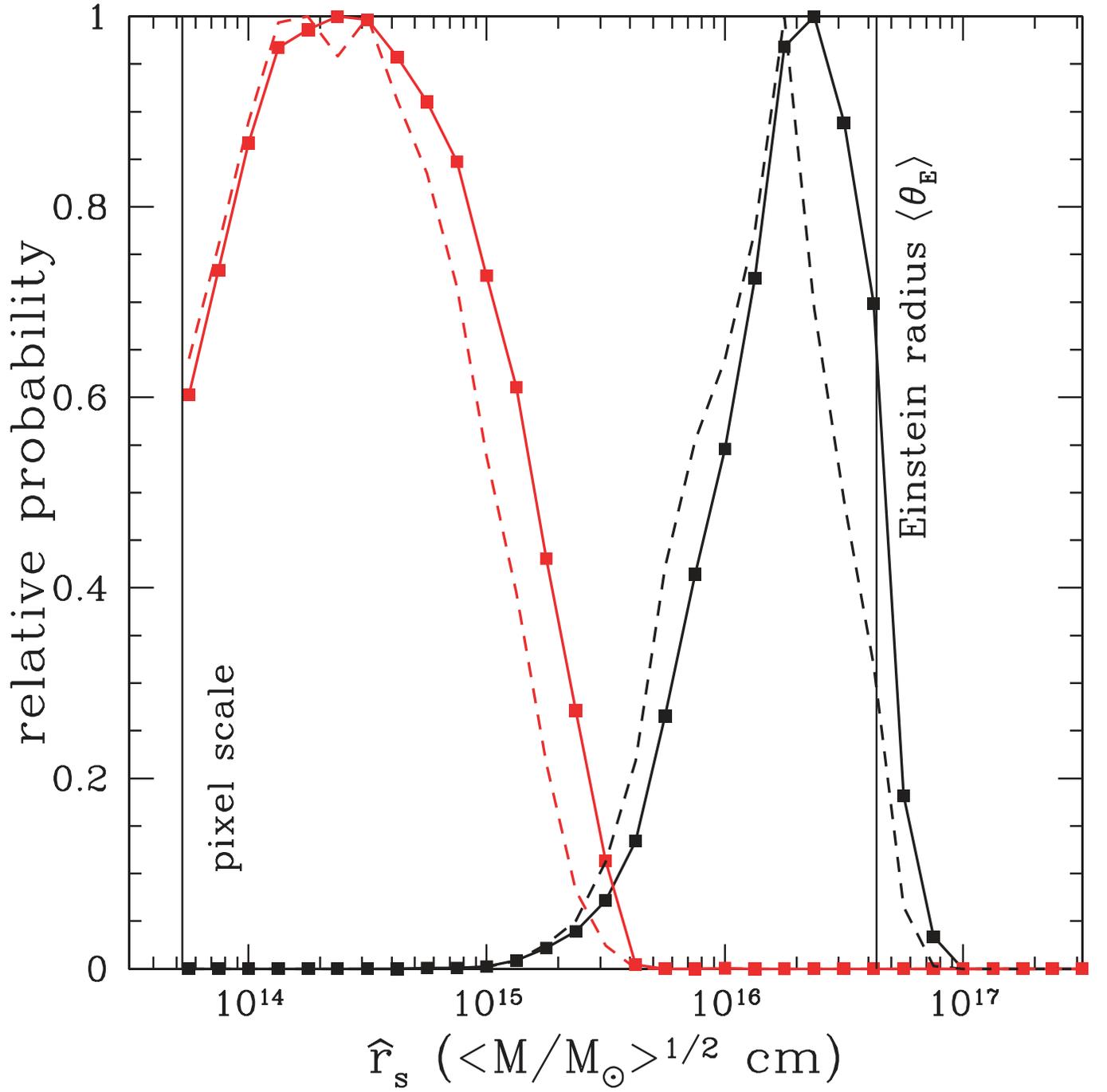}}
\caption{The probability density of the scaled source sizes $\hat{R}_\lambda$ in
  Einstein units of $\langle M/M_\odot \rangle^{1/2}$~cm.  Here the pixel scale
  and the mean Einstein radius correspond directly to the magnification patterns.
  The solid curves use only the velocity prior while the dashed lines add the
  mass prior.  The mass prior eliminates low mass, high velocity, large source
  size solutions.  Given our single X-ray data point, the probability distribution
  for the X-ray source would converge to a non-zero constant if we extended the
  calculations to still smaller sources using higher resolution magnification
  patterns.
 \label{f8.eps}}
\end{figure*}

\begin{figure*}
\centerline{\includegraphics[width=18cm]{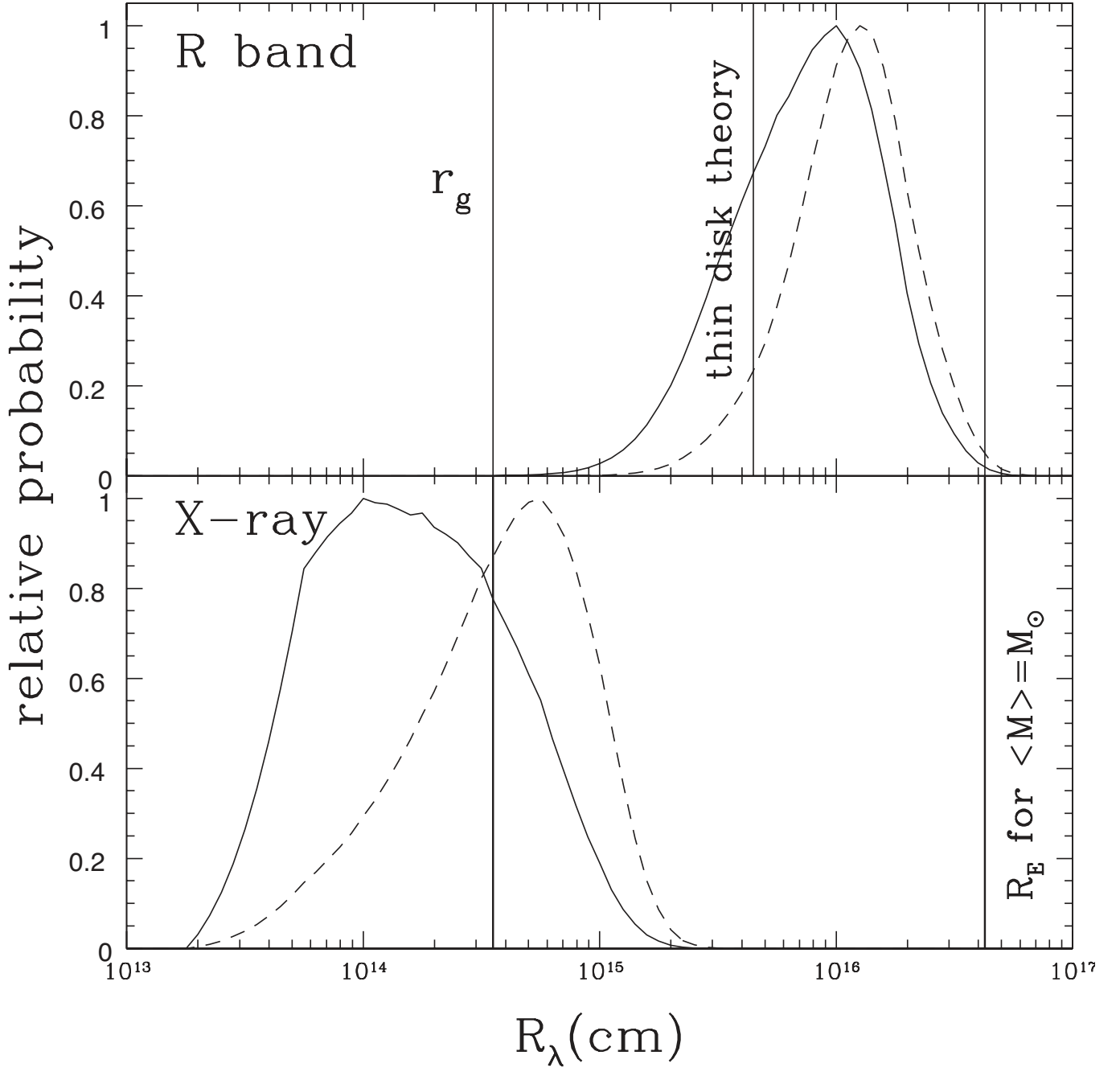}}
\caption{Disk size at 2000\AA\ (top) and X-ray (bottom) for \he, where the
  solid (dashed) curves assume logarithmic (linear) priors on the source size.
  We show the values of $R_\lambda$ for the assumed surface brightness profile
  so that the size ratios will be correct.  Physically, the disk size at 2000\AA\  can be 
  increased by a factor of $(\cos i)^{-1/2}$ for the inclination of the disk $i$, 
  and the X-ray size should be regarded as a measurement of the half-light radius
  with $R_{1/2}=2.44R_\lambda$.  Vertical lines mark the gravitational radius
  $r_g= GM_{BH}/c^2$ and accretion disk size given the wavelength, the estimated black hole mass,
  and Eddington limited accretion with an efficiency of $\eta=0.1$.  The final
  vertical line marks the Einstein radius for solar mass stars. 
 \label{f9.eps}}
\end{figure*}

\begin{figure*}
\centerline{\includegraphics[width=18cm]{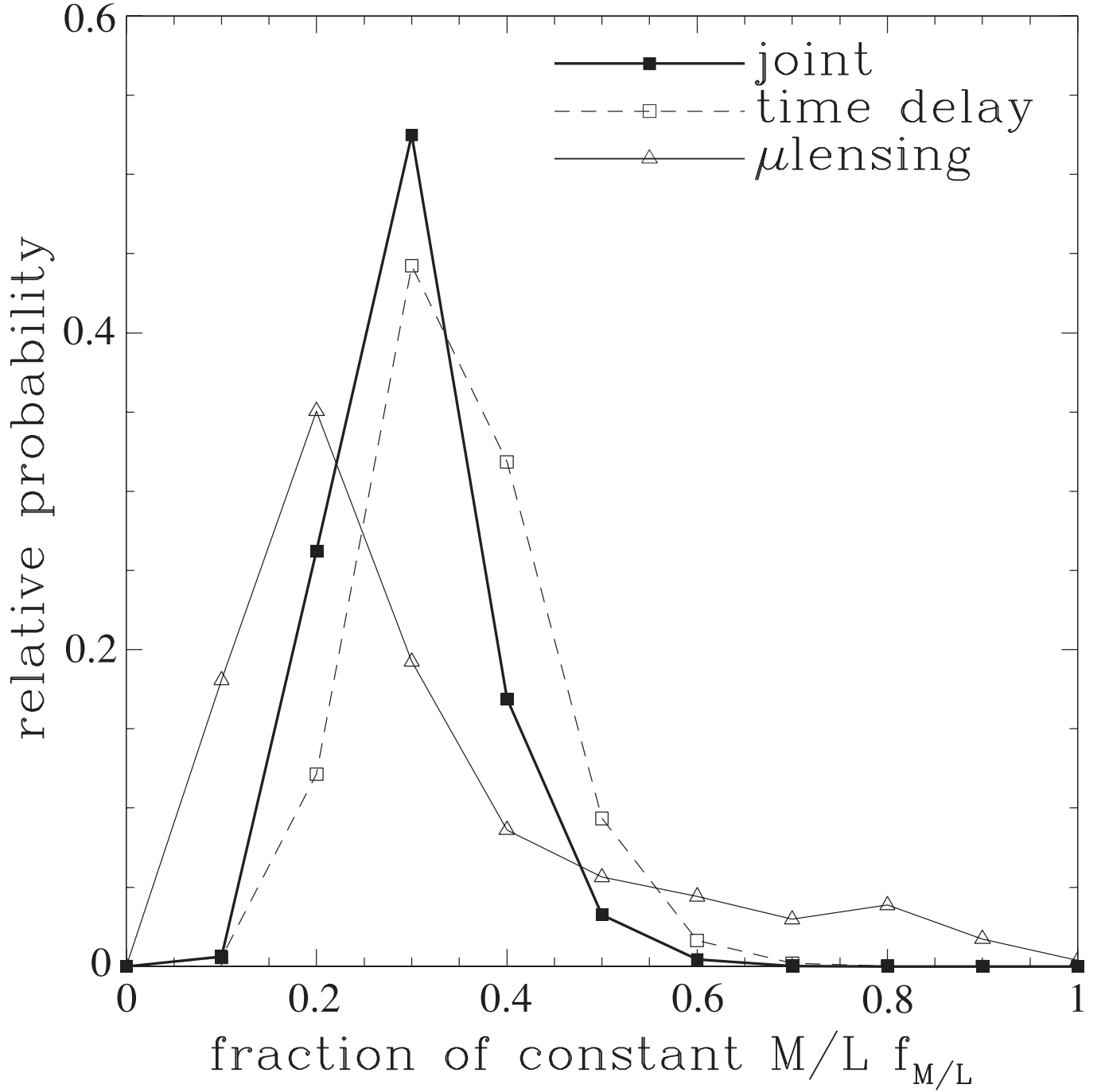}}
\caption{Constraints on the mass distribution.  The light solid, dashed and
  heavy solid curves show the constraints on the mass of the stellar 
  component of the lens mass model based on the microlensing models (heavy solid),
  the time delay measurements (dashed) and the combination of the two (light solid).  A 
  constant $M/L$ model has $f_{M/L}=1$ and a pure dark matter halo  
  model has $f_{M/L}=0$, with the stellar mass reduced from the 
  constant $M/L$ model in proportion to $f_{M/L}$
 \label{f10.eps}}
\end{figure*}

\begin{figure*}
\centerline{\includegraphics[width=18cm]{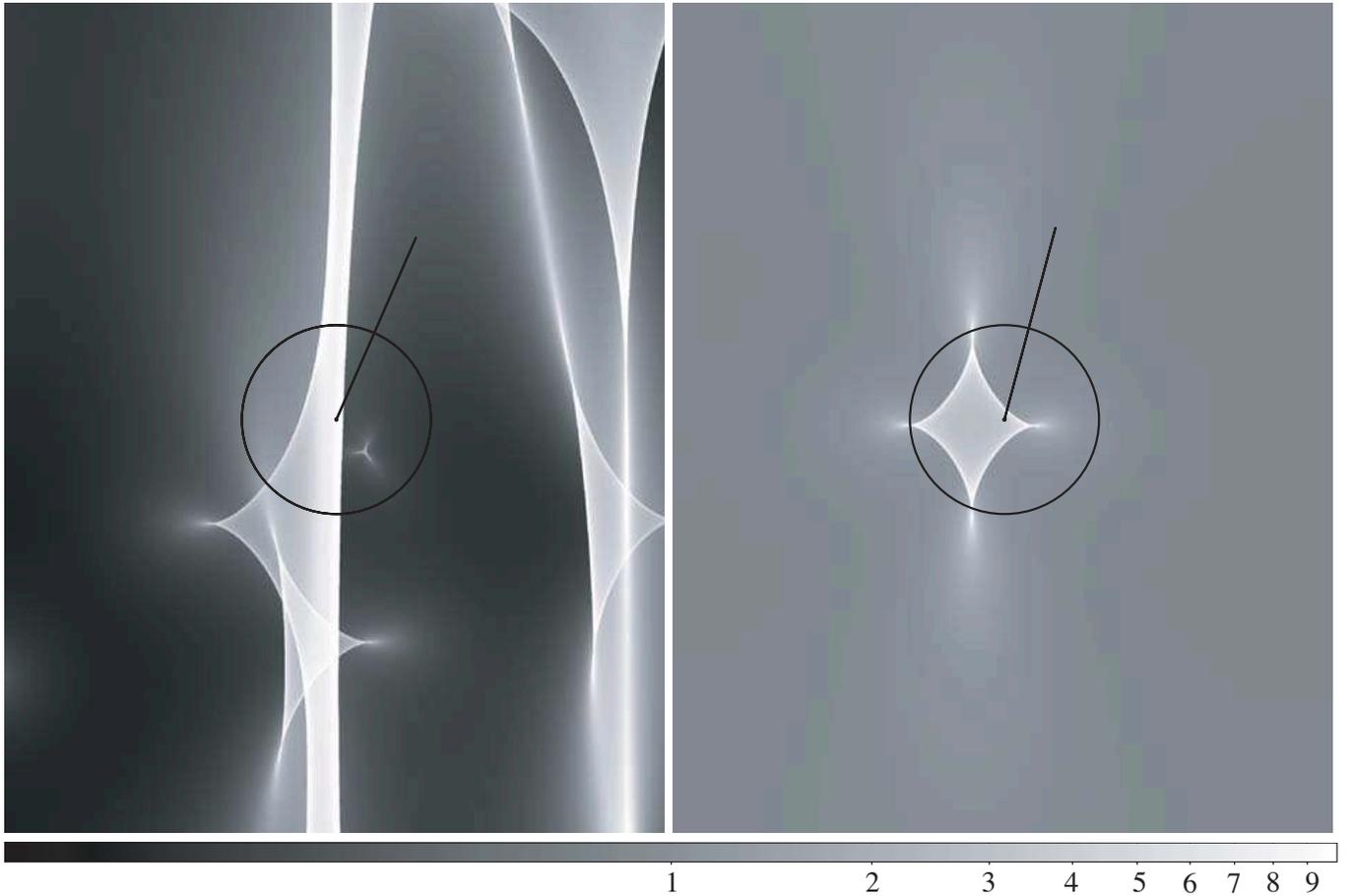}}
\caption{Simulated microlensing caustics for \he.  These show small
  sections of one pair of the $f_{M/L}=0.3$ magnification patterns for
  images A (left) and B (right) where higher magnifications are white
  and lower magnifications are black.  The lines show the path 
  followed by the source for the period covered by the $R$-band data starting 
  from the circled point.
  The large circle is the size of the UV (rest-frame) source ($\log \hat{R}_\lambda=16.2$)
  and the small (almost invisible) circle is the size of the X-ray
  source ($\log \hat{R}_\lambda=14.2$).
 \label{f11.eps}}
\end{figure*}

\begin{figure*}
\centerline{\includegraphics[width=18cm]{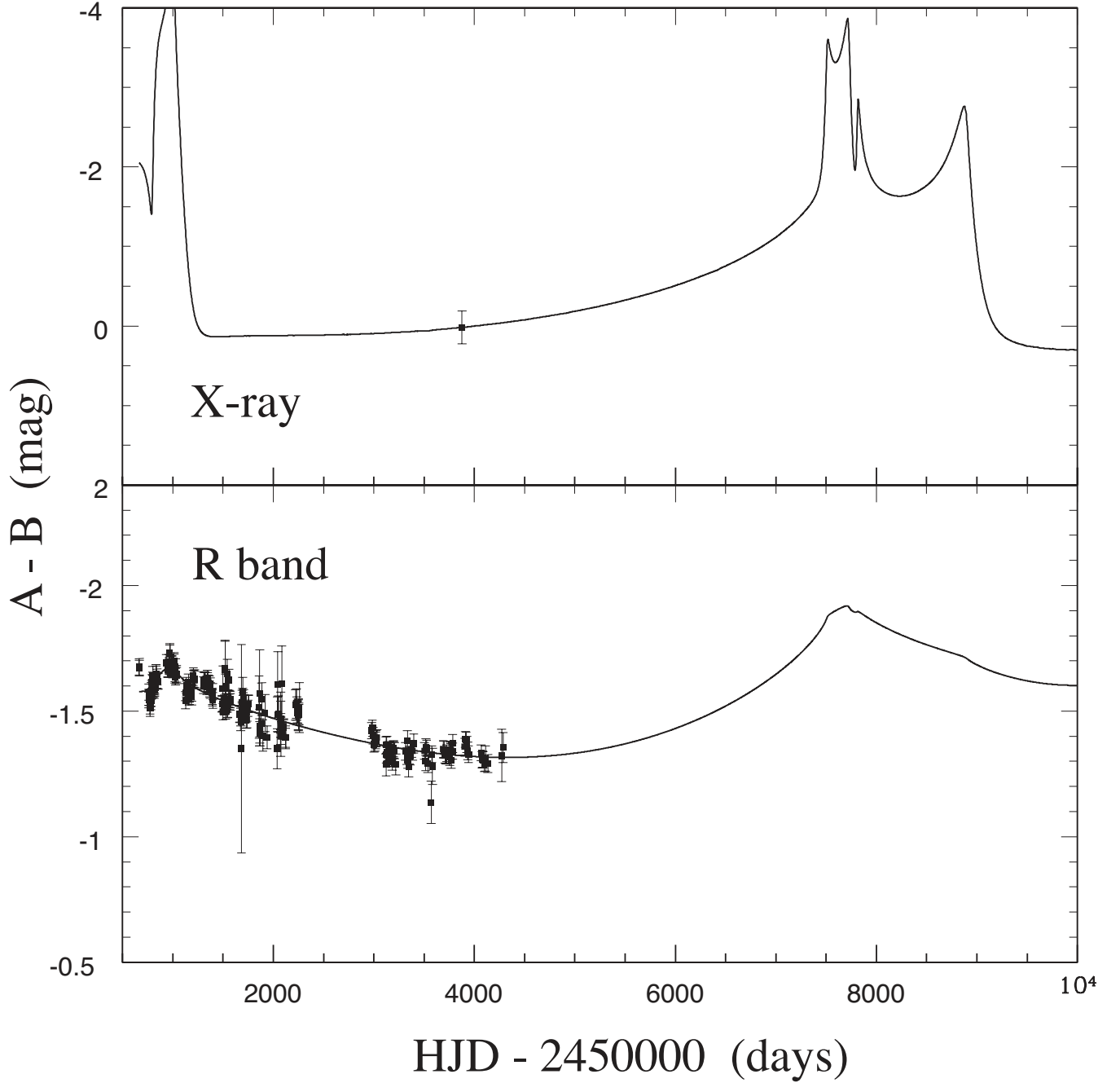}}
\caption{The $R$-band and X-ray magnitude differences for the trial shown in
   Fig.~11 as compared to the observations.  This trial was chosen because 
   it fit the data well and we found the origin of the initial peak as a
   pair of competing caustic crossings in the two images intriguing.  The
   dramatic differences in the vertical scales show that an X-ray light
   curve of this lens should be much more dramatic than the $R$-band.
 \label{f12.eps}}
\end{figure*}

\end{document}